\documentclass[twocolumn,preprintnumbers,amsmath,amssymb,prb,superscriptaddress]{revtex4-1}
\pdfoutput=1

\usepackage{graphicx}
\usepackage{float}
\usepackage{dcolumn}
\usepackage[tight]{subfigure}
\usepackage{amsmath}
\usepackage{verbatim}
\usepackage{units}
\usepackage[usenames,dvipsnames]{color}
\usepackage{bm}
\usepackage[normalem]{ulem}
\usepackage{enumerate} 

\usepackage{t1enc}
\usepackage[english]{babel}
\addto\captionsenglish{}

\renewcommand{\vec}[1]{\bm{#1} }

\newcommand{\bra}[1]{\langle #1|}
\newcommand{\ket}[1]{|#1 \rangle}

\newcommand{\Eq}[1]{Eq.~(\ref{#1})}

\newcommand{\Fig}[1]{Fig.\!~\ref{#1}}
\newcommand{\Figs}[1]{Figs.~\ref{#1}}

\usepackage[normalem]{ulem}

\usepackage{hyperref}
\hypersetup{colorlinks=true,linktoc=all,linkcolor=blue,citecolor=blue}
\usepackage{epstopdf}

\begin{document}


\title{Probing Transverse Magnetic Anisotropy by Electronic Transport\\ through a Single-Molecule Magnet}

\author{M. Misiorny}
\email{misiorny@amu.edu.pl}
\affiliation{Peter Gr{\"u}nberg Institut \& JARA, Forschungszentrum J{\"u}lich, 52425 J{\"u}lich,  Germany}
\affiliation{Faculty of Physics, Adam Mickiewicz University, 61-614 Pozna\'{n}, Poland}
\author{E. Burzur\'{\i}}
\email{E.BurzuriLinares@tudelft.nl}
\affiliation{Kavli Institute of Nanoscience, Delft University of Technology, 2600 GA, Delft, The Netherlands}
\author{R. Gaudenzi}
\affiliation{Kavli Institute of Nanoscience, Delft University of Technology, 2600 GA, Delft, The Netherlands}
\author{K. Park}
\affiliation{Department of Physics, Virginia Tech, Blacksburg, Virginia 24061, USA}
\author{M. Leijnse}
\affiliation{Solid State Physics and Nanometer Structure Consortium (nmC@LU), Lund University, Box 118, S-22100, Sweden}
\author{M. R. Wegewijs}
\affiliation{Peter Gr{\"u}nberg Institut \& JARA, Forschungszentrum J{\"u}lich, 52425 J{\"u}lich,  Germany}
\affiliation{Institute for Theory of Statistical Physics, RWTH Aachen, 52056 Aachen,  Germany}
\author{J. Paaske}
\affiliation{Center for Quantum Devices, Niels Bohr Institute, University of Copenhagen, 2100 Copenhagen, Denmark}
\author{A. Cornia}
\affiliation{Department of Chemical and Geological Sciences and INSTM, University of Modena and Reggio Emilia, via G. Campi 183, I-41125 Modena, Italy}
\author{H. S. J. van der Zant}
\affiliation{Kavli Institute of Nanoscience, Delft University of Technology, 2600 GA, Delft, The Netherlands}

\date{\today}



\begin{abstract}

By means of electronic transport, we study the transverse magnetic anisotropy of an individual Fe$_4$ single-molecule magnet (SMM) embedded in a three-terminal junction.
In particular, we   determine in situ the transverse anisotropy of the molecule from the pronounced intensity modulations of the linear conductance, which are observed as a function of applied magnetic field.
The proposed technique works at temperatures exceeding the energy scale of the tunnel splittings of the SMM.
We deduce that the transverse anisotropy for a single Fe$_4$ molecule captured in a junction is substantially larger than the bulk value.

\end{abstract}

\maketitle

\section{ Introduction}
Single-molecule magnets (SMMs)~\cite{Gatteschi_book}
have been proposed as candidates for applications in molecular spintronics.~\cite{Bogani_NatureMater.7/2008,Grose_NatureMater.7/2008,
Parks_Science328/2010,Zyazin_NanoLett.10/2010,Urdampilleta_NatureMater.10/2011,
Vincent_Nature488/2012}
Especially enticing is the prospect of using an individual SMM as a base component of a spintronic circuit which would be capable of storing \cite{Mannini_NatureMater.8/2009} or processing \cite{Leuenberger_Nature410/2001,Troiani_Chem.Soc.Rev.40/2011,Vincent_Nature488/2012,Thiele_Science344/2014} classical and quantum information.
In general, the essential prerequisite for this is a magnetic bistability which in
SMMs
stems from
a large molecular spin and a strong \emph{easy-axis}
magnetic anisotropy, given by a parameter $D$.
This  tends to fix the spin along an axis determined by the molecular structure,
without favoring any specific direction along this axis.
In consequence, an energy barrier $\sim \!\!DS^2$ protects the
spin of the molecule
against reversal between the two
opposite,
energetically degenerate  orientations.
From this point of view, detection of the additional \emph{transverse} magnetic anisotropy, characterized by the parameter
$E>0$
in the Hamiltonian
    $
    \hat{\mathcal{H}}
    =
    -D\hat{S}_z^2
    +
     E (\hat{S}_x^2-\hat{S}_y^2)
    $,
is crucially important. Such transverse anisotropy can impair the bistability by opening under-barrier quantum tunneling channels for spin reversal.~\cite{Gatteschi_book,Gatteschi_Angew.Chem.Int.Ed.42/2003,
Misiorny_Phys.Rev.Lett.111/2013}
These quantum tunneling processes are also of fundamental interest since the spin-dynamics displays \mbox{pronounced geometric or Berry-phase effects.~\cite{Wernsdorfer_Science284/1999,Romeike_Phys.Rev.Lett.96/2006_196601,
Romeike_Phys.Rev.Lett.96/2006_196805,Leuenberger_Phys.Rev.Lett.97/2006,
Gonzalez_Phys.Rev.Lett.98/2007,Burzuri_Phys.Rev.Lett.111/2013}}

Hitherto, most techniques aiming to extract the transverse anisotropy parameter $E$ are based on the detection of the tunnel splittings it induces, which display a characteristic magnetic field dependence.~\cite{Gatteschi_book,Gatteschi_Angew.Chem.Int.Ed.42/2003} The major challenge for all such approaches is that these splittings are complicated functions of $E$, and even more, the splitting for high-spin states and low magnetic fields are smaller than the parameter $E$ itself by several orders of magnitude.
Using Landau-Zener spectroscopy the tunnel splittings have been accurately determined in bulk Fe$_8$ by measuring their pronounced Berry-phase oscillations.~\cite{Wernsdorfer_Science284/1999}
Also in bulk crystals and solutions of SMMs the parameter $E$  has been established by different methods,
such as high-frequency electron paramagnetic resonance~\cite{Accorsi_J.Am.Chem.Soc.128/2006,Mannini_Nature468/2010} and inelastic neutron scattering.~\cite{Carretta_Phys.Rev.B70/2004}
These methods, however, probe large assemblies of molecules, and thus are not designed for investigating the magnetic properties of an individual SMM. As a result, little is known about the transverse anisotropy of individual SMMs in spintronic devices.

In this paper we propose an approach for extracting the parameter $E$ of a single molecule by employing electronic transport measurements.
We study a Fe$_4$ SMM captured in a gateable junction
(for details see App.~\ref{sec:Experiment})
---a geometry close to envisaged device structures--- which is a unique tool for addressing the spin in different redox states of a molecule.~\cite{Zyazin_NanoLett.10/2010}
We show that, as a consequence of the mixing of the spin eigenstates of the SMM, the transverse anisotropy significantly manifests itself in transport.  In particular, we predict and experimentally observe characteristic variations of the Coulomb peak \emph{amplitude} with the magnetic field  from which the parameter $E$ can be estimated.
Importantly, the method proposed here works at temperatures and electron tunnel broadenings $\Gamma$ exceeding $E$ by many orders of magnitude, while $E$ in its turn much exceeds the tunnel splittings.

\section{Three-terminal SMM junctions}
\begin{figure}[t]
  \includegraphics[width=0.99\columnwidth]{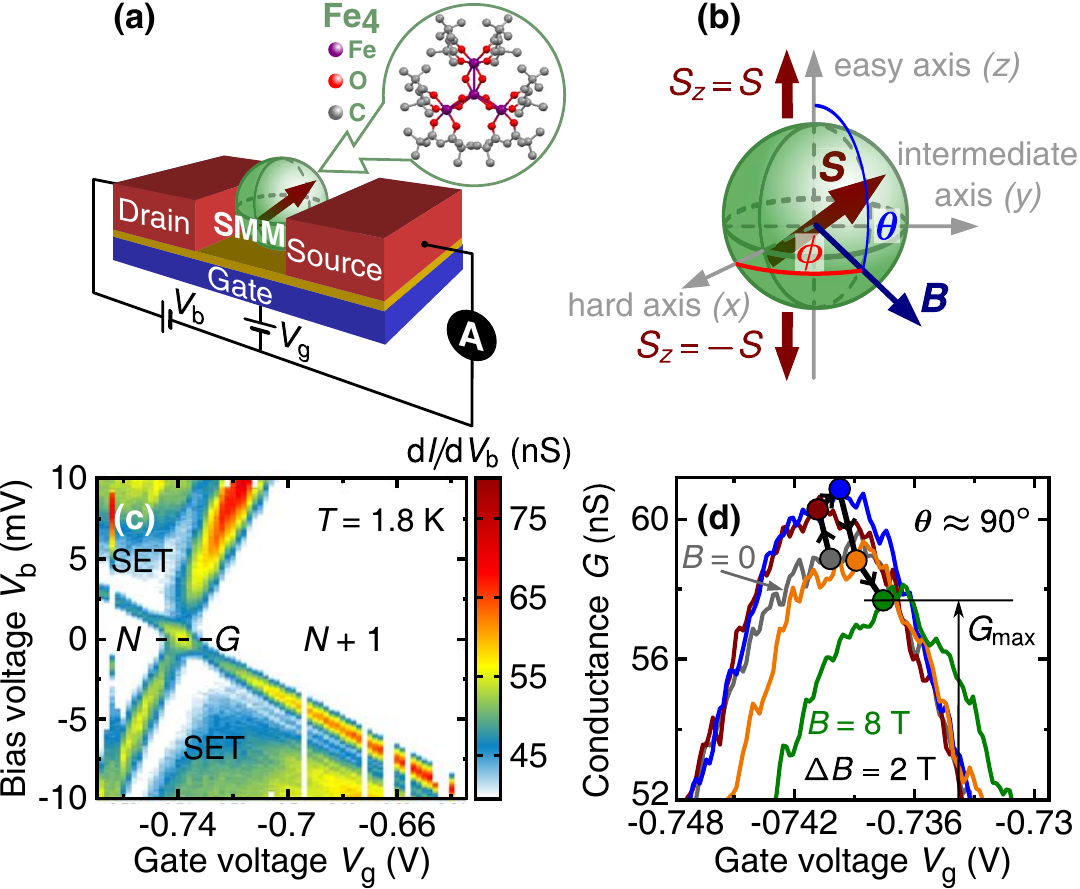}
  \caption{
  (color online)
  (a) Schematic depiction of a molecular three-terminal transistor with a single Fe$_4$ SMM bridging the junction.
  (b) Spatial orientation of an external magnetic field with respect to the principal axes set by the magnetic anisotropy of an SMM.
  (c) Differential-conductance map, $\text{d}I/\text{d}V_\text{b}$, measured as a function of gate $V_\text{g}$ and bias $V_\text{b}$ voltages showing two charge states $N$ (neutral) and $N+1$ (charged)
  for Sample A.
  (d) Representative Coulomb peaks [corresponding to linear conductance $G\equiv \text{d}I/\text{d}V_\text{b}|_{V_\text{b}=0}$ -- e.g., marked by  dashed line in (c)]  measured at different values of the external magnetic field $B$.
  The bold arrowed lines and color dots serve  as a guide for eyes to indicate the non-monotonic change in the Coulomb peak height.
  }
\label{Fig_1}
\vspace*{-\baselineskip}
\end{figure}

A scheme of a three-terminal SMM junction is shown in \Fig{Fig_1}(a).
An SMM bridges the source and drain gold electrodes. An underlying aluminium electrode separated by a few nanometers of aluminium oxide allows for electrical gating of the molecule and, thus, accessing different redox states,
see also App.~\ref{sec:Three-terminal_junctions}.
The chip containing the junctions is mounted on a piezo-driven rotator that enables
to change \emph{in situ} the orientation between the external magnetic field $\vec{B}$ and the magnetic anisotropy axes of the molecule, which is characterized by angles $\theta$ and~$\phi$ as illustrated in \Fig{Fig_1}(b). All the measurements are performed at $T=1.8$~K.

The differential conductance plotted in \Fig{Fig_1}(c) shows the standard signatures of sequential electron tunneling (SET) through a molecule with two competing charge states
tuned by a gate voltage.~\cite{Ferry_book}
Strong high-conductance resonance lines separate adjacent charge-stable Coulomb blockade regions, labeled $N$ and $N+1$, from the SET regions where transport is possible.
Importantly, several fingerprint features of the stable Fe$_4$ SMM can be identified:
(i) high charging energies expected for an individual molecule;
(ii) a strong SET excitation at approximately 4.8 meV,~\cite{Zyazin_NanoLett.10/2010} specific to Fe$_4$ as it corresponds to the predicted transition energy between the ground ($S_N=5$) and the first-excited ($S_N=4$) spin multiplets for the neutral molecule;~\cite{Accorsi_J.Am.Chem.Soc.128/2006}
(iii) a non-linear shift of the degeneracy peak in the presence of magnetic field as described by gate-voltage spectroscopy
(for details see Ref.~[\onlinecite{Burzuri_Phys.Rev.Lett.109/2012}] and App.~\ref{sec:Gate-voltage_position}).
Moreover, depending on the strength of tunnel coupling~$\Gamma$, split Kondo zero-bias anomalies in Coulomb blockade regimes of \emph{subsequent} charge states can be observed, which show the zero-field splitting (ZFS) at the values expected for the Fe$_4$ SMM.~\cite{Zyazin_NanoLett.10/2010,Zyazin_Synt.Met.161/2011}
These features also indicate that the molecule is in an intermediate coupling regime with the electrodes,
with its upper-limit estimated to be
$\Gamma$=1.6 meV -- obtained from the full width at half maximum of the
crossing (degeneracy) point of the Coulomb edges at zero bias, the Coulomb peak,
for further discussion see App.~\ref{sec:Gate-voltage_position}.

\section{Gate-voltage `position' spectroscopy}

In a magnetic field the position of the
Coulomb peak (CP) depends both on the \emph{magnitude} and the \emph{orientation} of an external magnetic field $\vec{B}$.~\cite{Burzuri_Phys.Rev.Lett.109/2012}
 In short, the CP marks the transition between the ground states of two spin multiplets, with spin values $S_N$ and $S_{N+1}$, for the two neighboring charge states. The energy difference between these states is then a function of $\vec{B}$, and in particular, it translates into a shift of the linear response degeneracy point in $V_\text{g}$, as shown in \Fig{Fig_1}(d).
From such a shift one can infer that the ground spin-multiplets of the two charge states evolve differently in the applied field; therefore, the shift provides information about the magnetic properties of the system.
For example, in simple quantum dots the shift corresponds just to the \emph{linear} Zeeman effect which is \emph{isotropic}.~\cite{Hanson_Rev.Mod.Phys.79/2007}
On the other hand, for magnetically anisotropic molecules, like the SMMs discussed here, not only does the CP shift depend on the relative sample-field orientation, allowing us to extract the value of the angle $\theta$, but it also  provides information about the uniaxial magnetic anisotropy~($D$).~\cite{Burzuri_Phys.Rev.Lett.109/2012}
However,
the gate-voltage position of the peak, determined by the low-energy spectrum, is insensitive to the small tunnel splitting corrections induced by the transverse magnetic anisotropy.
Below we show that
information about the transverse magnetic anisotropy~($E$) can instead be inferred from a nonmonotonic dependence of the peak \emph{amplitude} $G_\text{max}$, such as in \Fig{Fig_1}(d),
which relies on transition probabilities between different spin~states.
We have measured around 200 junctions of which 17 showed clear molecular signatures. From those, 9 samples displayed a clear Coulomb peak suitable to perform gate spectroscopy and a magnetic field modulation of $G_\text{max}$.
Further discussion of statistics together with differential-conductance maps for several devices are presented in App.~\ref{sec:Statistics}.

\begin{figure}[t]
  \includegraphics[width=0.99\columnwidth]{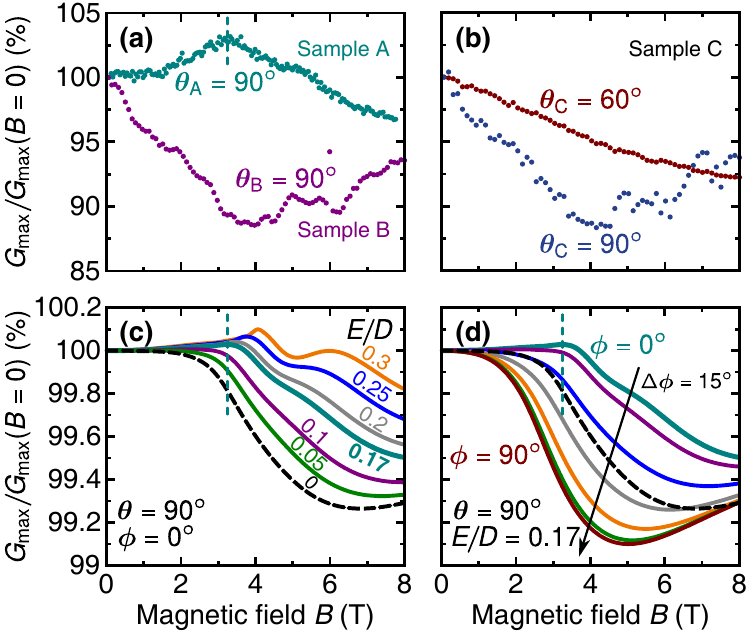}
  \caption{
  (color online)
  Signatures of transverse magnetic anisotropy in electronic transport at $T=1.8$ K:
  (a) Dependence of the Coulomb peak (CP) height $G_\text{max}$  [i.e., the maximal value of $G$, cf. \Fig{Fig_1}(d)] on magnetic field $B$ shown for two different samples where the orientation of the magnetic field lies in the hard plane ($\theta=90^\circ$).
  (b) Analogous to (a) for a single sample, except that now $\theta$ is varied and $\phi$ is unknown.
  Note that the evolution of the CP position in magnetic field, and not $G_\text{max}$, was previously analyzed in Ref.~[\onlinecite{Burzuri_Phys.Rev.Lett.109/2012}] for samples A and C.
  \emph{Bottom panels:}
  Theoretical predictions for evolution of  the CP height with magnetic field $B$ kept in the hard plane:  (c) for  indicated values of $E/D$ and $\phi=0^\circ$, whereas in (d)  for several  angles $\phi$  and the fixed value of $E/D$ estimated from~(a).
  Bold dashed lines represent the case of $E/D=0$ for $\phi=0^\circ$ (c) and $\phi=90^\circ$~(d).
  Notice that
  the shape of $G_\text{max}$ for $E/D=0$ is independent of $\phi$ due to the rotational symmetry around the molecule's easy axis.
  }
\label{Fig_2}
\vspace*{-\baselineskip}
\end{figure}

In \Fig{Fig_2}(a) the amplitude $G_\text{max}$ of the Coulomb peak, normalized to its value at $B=0$, is plotted as function of $B$ for two different samples.
For both samples, the gate-voltage analysis of the peak position allows us to conclude that the magnetic field lies in the hard plane ($\theta\approx90^\circ$),
see App.~\ref{sec:Gate-voltage_position}.
Interestingly, $G_\text{max}(B)$ for the two samples exhibits a significantly different behavior.
If only uniaxial magnetic anisotropy was present ($E=0$),
the transport properties of the molecule would be left unaffected upon rotation of the field in the hard plane.
On the contrary, for $E\neq0$
this rotational symmetry is broken. The dissimilar behavior of the {amplitude} $G_\text{max}$ as observed in \Fig{Fig_2}(a) is therefore attributed to different values of the angle $\phi$ in the presence of a non-zero $E$.
Similar curve shapes have been observed in additional samples,
as shown in Fig.~\ref{Fig_9}.
Although the values of $E$ for bulk samples/monolayers of SMMs are typically small (for Fe$_4$ $E/D\lesssim0.07$)~\cite{Gregoli_Chem.Eur.J.15/2009,Mannini_Nature468/2010}, the  linear conductance through a molecule appears to be measurably influenced by it.
A similar change in the field-evolution of $G_\text{max}$ is also observed in a single sample C, shown in \Fig{Fig_2}(b),  by rotating the sample holder relative to the magnetic field.

\section{Theory and discussion}
In order to understand how the transverse magnetic anisotropy could
qualitatively affect the linear conductance through an SMM (i.e., the CP amplitude), while hardly influencing its gate-voltage position,
 we use a minimal molecular quantum-dot model based on two giant-spin Hamiltonians,~\cite{Gatteschi_book}
    \begin{equation}\label{eq:H_SMM}
    \hat{\mathcal{H}}_\text{SMM}
    =
    \!\!
    \sum_{n=N,N+1}
    \!\!
    \big[
    \hat{\mathcal{H}}_n^{}
    +
    \hat{\mathcal{H}}_n^\text{Z}
    \big]
    ,
    \end{equation}
one for each charge state.
Here, $\hat{\mathcal{H}}_n$ accounts for the magnetic anisotropy of the SMM in the $n$th charge~state,
    \begin{equation}\label{eq:H_N}
    \hat{\mathcal{H}}_n
    =
    - D_n \big(\hat{S}_n^{z} \big)^{\!2}
    +
    E_n\Big[\big(\hat{S}_n^x \big)^{\!2}-\big(\hat{S}_n^y \big)^{\!2}\Big]
    ,
    \end{equation}
with the first/second term representing the uniaxial/transverse magnetic anisotropy, and $\hat{\mathcal{H}}_n^\text{Z}=g\mu_\text{B}\vec{B}\cdot\hat{\vec{S}}_n$ is the Zeeman term ($g\approx2$).
We combine this with a master equation description of the SET transport to nonmagnetic electrodes with tunnel coupling $\Gamma$.~\cite{Romeike_Phys.Rev.Lett.96/2006_196805,
Timm_Phys.Rev.B73/2006,Misiorny_Phys.Rev.B79/2009}
The essential steps of this approach are provided in App.~\ref{sec:Transport}.
The appearance of a clear CP in the experiment restricts $S_{N+1}=S_N\pm1/2$ (otherwise spin-blockade would be seen)~\cite{Zyazin_NanoLett.10/2010}.
For the Fe$_4$ SMM we can estimate $S_N=5$ and $D_N\equiv D\approx 56$ $\mu$eV for the neutral state, whereas from the CP position dependence
we obtain $S_{N+1}=9/2$,
and  fix
$D_{N+1} \approx 1.2 D = 68$ $\mu$eV with approximately collinear easy axes for both charge states,
all in agreement with previous measurements,~\cite{Burzuri_Phys.Rev.Lett.109/2012}
see also App.~\ref{sec:Gate-voltage_position}.
We assume that upon charging only the overall energy scale of the magnetic anisotropy changes, i.e., $E_N/D_N \approx E_{N+1}/D_{N+1}$, leaving just a single parameter~$E_N=E$ for the transverse anisotropy.

In \Fig{Fig_2}(c) we plot the calculated CP amplitude  $G_\text{max}$ for $\theta=90^\circ$ and $\phi=0^\circ$ as a function of the applied field $B$.
Surprisingly, the calculations reveal that a non-zero value of $E$ significantly influences the current through the molecule.
By adjusting the parameter $E/D$, qualitative agreement with the measured amplitude variation is obtained for sample A when
$E/D\approx0.15-0.2$.
The dissimilar behavior of $G_\text{max}$  between samples A and B is then qualitatively reproduced when assuming strongly differing values of the angle $\phi$
as shown in  \Fig{Fig_2}(d).
From the shape of the curves we estimate the value of $\phi$ to be $\phi_A\approx0^\circ$ for sample A and $\phi_B\approx90^\circ$ for sample B. Note that the minimum of $G_\text{max}$ for $\phi=90^\circ$ appears in \Fig{Fig_2}(d) at a somewhat larger $B$ field value than for sample B which signifies larger $E/D$,
cf. Figs.~\ref{Fig_13}-\ref{Fig_15}.
Therefore, combining the information from \Figs{Fig_2}(c)-(d), the CP amplitude could be used to estimate the values of $E$ and $\phi$.
The obtained rough estimate $E/D\approx 0.17$ is larger than the values reported for bulk samples,~\cite{Gregoli_Chem.Eur.J.15/2009} as also suggested by XMCD experiments on Fe$_4$ monolayers deposited on gold.~\cite{Mannini_Nature468/2010}

\begin{figure}[t]
  \includegraphics[width=0.99\columnwidth]{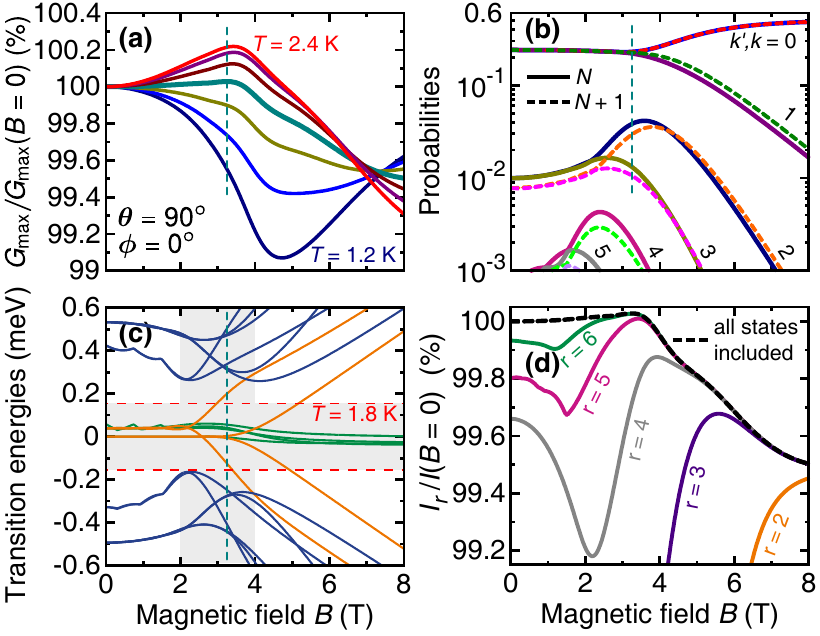}
  \caption{
  (color online)
  Theoretical analysis of transport~for fixed $D=56$ $\mu$eV and $E/D=0.17$
  and $\vec{B}$ along the hard axis
  ($\theta=90^\circ$ and $\phi=0^\circ$):
  (a) Conductance $G_\text{max}(B)$ traces for various temperatures over the range $1.2-2.4$ K at intervals of 0.2 K.
  (b) Occupation probabilities for several lowest-energy states in
  the spin multiplets for $N$ and $N+1$ at $T=1.8$ K. Here, $k^\prime$ ($k$)
  labels the states in order of increasing energy
 for $N$ ($N+1$), with $k^\prime=0$ ($k=0$) denoting the ground state.
  (c) Relevant transition energies $\varepsilon_{N+1}^k-\varepsilon_N^{k^\prime}$ for $k,k^\prime\leqslant4$ determining the SET processes at the Coulomb resonance (note that $\varepsilon_{N+1}^0=\varepsilon_N^0$ is restored for each $B$ by tuning $V_\text{g}$). Different colors of lines  are used to distinguish groups of transitions with respect to possible combinations of indices $k$ and $k^\prime$ (see the main text).
  For the association of these lines with specific transitions as well as the energies of individual levels
  see Fig.~\ref{Fig_4}.
  (d)
  Evolution of the current vs. magnetic field
  at $T=1.8$~K
  calculated  by including a restricted number of states per spin-multiplet
  up
  to~$r$, where  $r=k_\text{max}^\prime+1=k_\text{max}+1$, showing that for small $r$ significant deviations are found compared to the calculation involving all the states (dashed line), used in all other plots.
  For a precise definition of the current $I_r$ see App.~\ref{sec:Transport}.
  }
\label{Fig_3}
\end{figure}

To gain deeper insight into the mechanism leading to a modulation of $G_\text{max}$ we analyze in \Fig{Fig_3}(a) how the calculated $B$-traces of the CP amplitude evolve with temperature.
The appearance of a maximum at around $B=3.25$~T (marked by the vertical dashed line) and its enhancement with increasing temperature suggests that this feature is build up from contributions of many excited states of the SMM.
This is indeed confirmed by inspection of the evolution of the occupation probabilities shown in \Fig{Fig_3}(b) for the experimental temperature $T=1.8$ K.
To obtain this figure we first find the eigenstates of $\hat{\mathcal{H}}_n$, given by Eq.~(\ref{eq:H_N}). For $n=N,N+1$ we obtain two sets of eigenspectra, $\{\varepsilon_N^{k^\prime}\}$ and $\{\varepsilon_{N+1}^k\}$. Here, $k^\prime$ and $k$
label
 the states in order of increasing energy, starting from $k^\prime=0$ ($k=0$) for the neutral (charged) \emph{ground} state.
Using these energies and states, we calculate the probabilities from the master equation.
One should note that the energies (not shown) and occupation probabilities of corresponding states ($k=k^\prime$) for different charge  are very similar.
From \Fig{Fig_3}(b), however,
it is not clear which of the maxima of the probabilities is responsible for the maximum of
 the $G_\text{max}(B)$ curves, indicated by the vertical dashed line.

\begin{figure}[t]
   \includegraphics[scale=0.85]{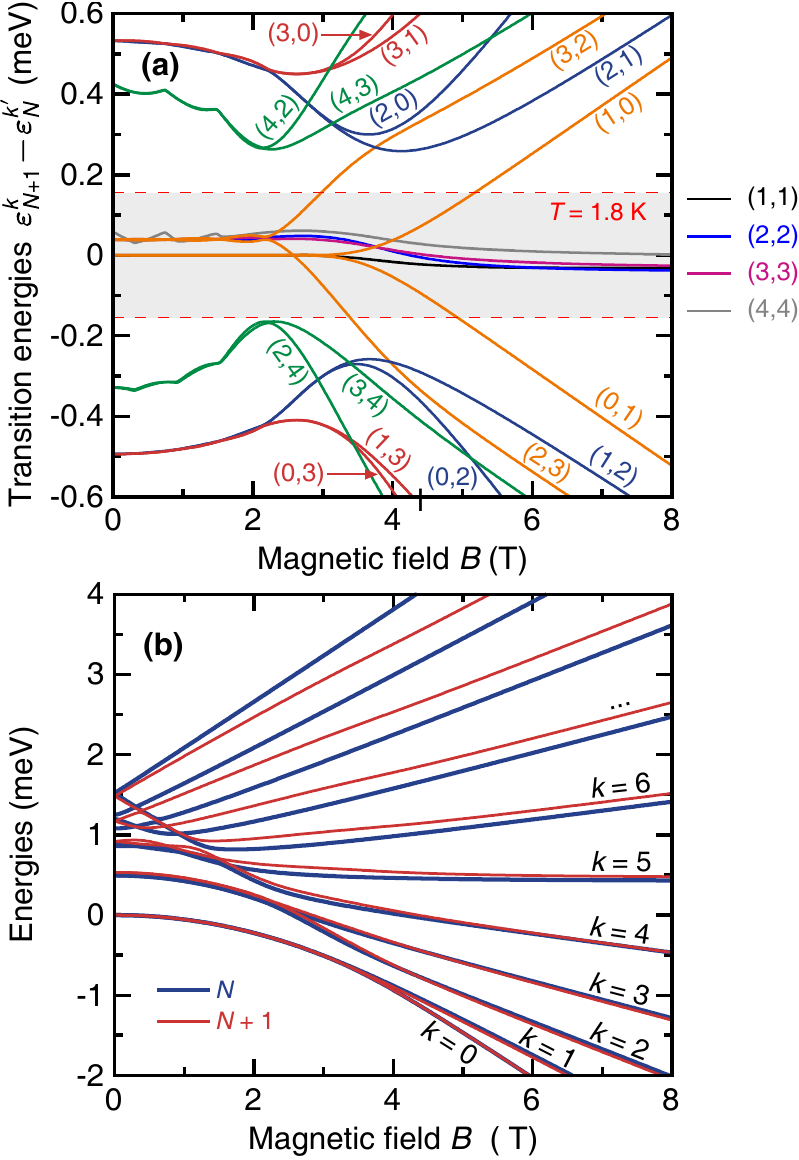}
    \caption{
    (color online)
    Panel~(a) is identical to Fig.~~\ref{Fig_3}(c),  but now for each transition-energy line we specify the
    initial and final states, with respective energies $\varepsilon_{N}^{k^\prime}$ and $\varepsilon_{N+1}^k$,
    between which the transition occurs.
    Recall that $k$ is an index which numbers states in a given spin multiplet with respect to energy, with $k=0$ denoting the ground state. Moreover, by labelling the lines with  $(k,k^\prime)$ we mean that $k$ refers to the final state of a charged SMM ($N+1$) whereas $k^\prime$ represents the initial state of a neutral SMM ($N$).
    We note that information shown in (a) cannot be readily seen from energies $\varepsilon_n^k$  ($n=N,N+1$) of the individual levels, which for the completeness of the present discussion are plotted in
    (b).
    Observe that since energies in (b) are calculated at the Coulomb resonance, the curves for $k=0$ overlap.
  }
  \label{Fig_4}
\end{figure}

Instead, to understand the $G_\text{max}(B)$ dependence in \Fig{Fig_3}(a) one has to consider
the transition energies
$\varepsilon_{N+1}^k-\varepsilon_N^{k^\prime}$
between levels of different charge states.
This is demonstrated in
 \Fig{Fig_3}(c), where the horizontal dashed lines represent the available thermal energy.
The  transition energies fall into three generic groups:
(i) low-energy transitions ($k=k^\prime$ -- green lines);
(ii) transitions of low energy for small $B$ but high energy for large $B$ ($k,k^\prime=0,1$ or $k,k^\prime=2,3$ -- orange lines);
(iii) high-energy
transitions
 (remaining $k$ and $k^\prime$ pairs -- blue lines).
Importantly, the temperatures used in \Fig{Fig_3}(a) lie just below the group of transition-energy curves exhibiting a minimum at finite magnetic fields roughly between 2-4 T (blue curves in \Fig{Fig_3}(c)). As the magnetic field is augmented from zero, these curves thus initially approach the thermal energy (horizontal dashed lines) before moving away at higher fields towards their high-field asymptotes. This leads to an enhancement of $G_\text{max}$ for $B\lesssim3.25$~T, followed by a steady decrease, i.e., the characteristic non-monotonic behavior
 experimentally observed in \Fig{Fig_2}(a).
We emphasize that the above mechanism does not constitute a purely spectroscopic method: the current and probabilities depend on both  the \emph{energies} and \emph{quantum states}, which determine the tunnel rates.
The importance of including many excited states in the calculation is quantified in~\Fig{Fig_3}(d), where we show how the non-monotonic behavior can be strongly overestimated when including too few  excited states,
see also Figs.~\ref{Fig_13}-\ref{Fig_16}.
We note that some additional remarks regarding signatures of the transverse anisotropy parameter $E$ in the peak amplitude of $G_\text{max}$ are discussed in App.~\ref{sec:Signatures_E}.

Finally, worth of note is the larger-than-predicted modulation of the CP amplitude observed in the experiments. We briefly comment on the verifications to rule out some other contributions that could lead to such an amplification.  First, the master equation analysis was constrained to a weak tunnel-coupling $\Gamma$ as compared to temperature. We verified that higher-order tunnel processes that lead to broadening and inelastic tunneling do not increase the scale of the modulation of the  CP height. For this we employed a perturbative approach including next-to-leading tunneling processes~\cite{Leijnse_Phys.Rev.B78/2008} and non-perturbative numerical renormalization group (NRG) method.~\cite{Bulla_Rev.Mod.Phys.80/2008,Toth_Phys.Rev.B78/2008,Legeza_DMNRGmanual}
Second, we assumed symmetric tunnel-coupling of the SMM to both electrodes  with the same energy $\Gamma$. One can show that a junction asymmetry gives rise to an overall constant factor suppressing the conductance $G_\text{max}$. Thus, this cannot change its field dependence.
Third,
the addition of
higher-order magnetic anisotropy terms to
the SMM model,~\Eq{eq:H_N}, is also not likely to affect the magnitude of the modulation. We checked, for instance, the effect of the 4th order transverse anisotropy of the form
    $
    C_n\big[\big(\hat{S}_n^x \big)^{\!4}-\big(\hat{S}_n^y \big)^{\!4}\big]
    $,
for a range of values of the parameter $C_{N/N+1}$ for which this term competes with the 2nd order transverse term. We thus conclude that the intensity of the modulation may rely on some intrinsic amplification mechanism not captured by our model, i.e., going beyond the giant-spin model,~\cite{Wilson_Phys.Rev.B74/2006,Burzuri_Phys.Rev.Lett.111/2013} when considering a single electron interacting with the molecule.

\section{\label{sec:Fitting_procedure}Fitting procedure: how to find anisotropy parameters of a single molecule from its transport spectra}

We summarize here in a few steps how to determine magnetic anisotropy of an individual SMM (see Eqs.~(\ref{eq:H_SMM})-(\ref{eq:H_N}) and App.~\ref{sec:Model})
by exploiting the information contained both in the Coulomb peak \emph{position} as well as in the magnetic field evolution of its \emph{amplitude}. In particular, the method under discussion allows for finding both the magnetic anisotropy constants $D_n$ and $E_n$ in two charge states (i.e. for $n=N,N+1$) of an SMM, and the orientation of an external magnetic field relative to the molecule's principle axes, given by the angles $\theta$ and $\phi$.

\begin{figure*}[t!!!]
   \includegraphics[scale=0.8]{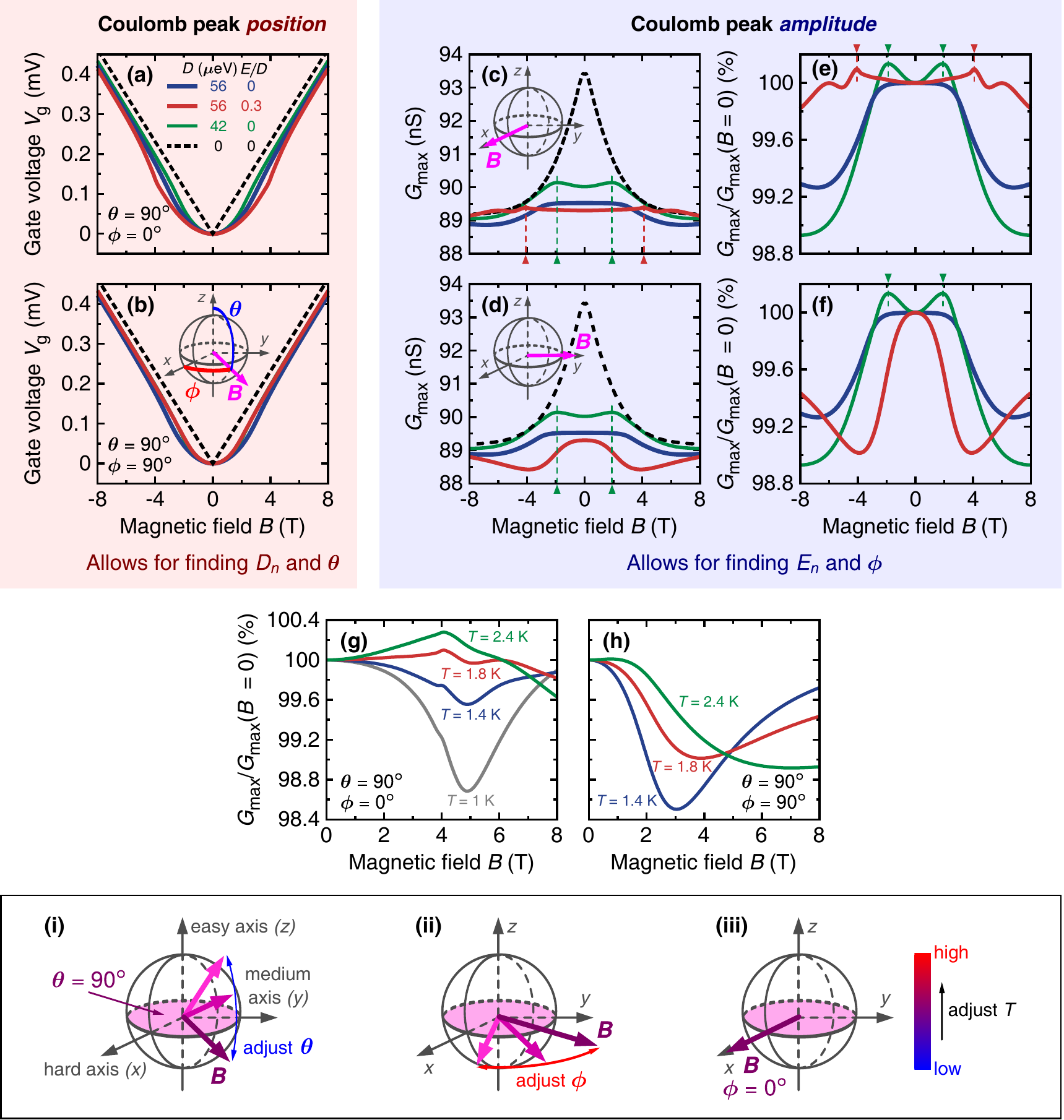}
    \caption{
    (color online)
    How to determine the transverse magnetic anisotropy constant $E$ of an individual SMM from its transport characteristics:
    The \emph{position} (a)-(b) and \emph{amplitude} (c)-(f) of the Coulomb peak are shown for  different values of the parameters $D$ and $E$  of the SMM model with $S_N=5$ and $S_{N+1}=9/2$ for $T=1.8$ K. Note that we employ the assumption for the Fe$_4$ molecule from the main text, that is $D=D_N=D_{N+1}/1.2$ and $E=E_N$ with $E_N/E_{N+1}=D_N/D_{N+1}$, and a relatively large value of $E/D$ (red lines) is used for clear illustration of the effects under discussion.
    In panels (a,c,e) the external magnetic field $\vec{B}$ is oriented along the SMM's \emph{hard} axis $x$ [see inset in (c)], whereas in panels (b,d,f) the field is parallel to the \emph{medium} axis $y$ [see inset in (d)].
    In panel (g) we present how temperature affects the occurrence of characteristic peaks associated with the presence of transverse magnetic anisotropy for $\vec{B}$ along the \emph{hard} axis $x$ -- for further details see \Fig{Fig_16}. To make the discussion complete, in panel (h) we show analogous dependencies but in the case when the field lies along the \emph{medium} axis $y$.
    Finally, the frame at the bottom contains a schematic summary of the procedure leading to estimation of $E$:
    (i) Using the analysis of the Coulomb peak \emph{position}, find $D_n$ and adjust the magnetic field $\vec{B}$ so that it is contained in the hard plane, i.e., the plane perpendicular to the easy axis $z$.
    (ii) Rotating systematically the magnetic field $\vec{B}$ in the hard plane, analyze the Coulomb peak \emph{amplitude} to find the direction of the molecule's hard axis.
    This will be characterized by occurrence of additional peaks in the amplitude, whose field-position allows for estimating $E_n$.
    (iii) If no local maxima in the amplitude can be seen, adjust (try increasing) the temperature.
  }
  \label{Fig_5}
\end{figure*}

\begin{enumerate}[(i)]
\item
Let us first consider only the Coulomb peak \emph{position}, shown in the left panel of \Fig{Fig_5}.
As explained in Ref.~[\onlinecite{Burzuri_Phys.Rev.Lett.109/2012}], by analyzing the position of the Coulomb peak one can immediately conclude whether a molecule captured in the junction exhibits magnetic anisotropy at all. If the molecule is \emph{spin-isotropic}, one observes a linear dependence on the magnetic field [see dashed line in \Figs{Fig_5}(a)-(b)] that reflects the linear Zeeman effect. On the other hand, if the molecule is \emph{spin-anisotropic}, this dependence becomes nonlinear, and the uniaxial magnetic anisotropy parameter $D_n$ together with the angle $\theta$ can be estimated from it.
This, in turn, permits for systematic adjustment of the magnetic field's orientation so that the field is kept perpendicular to the molecule's easy axis $z$, which corresponds to $\theta=90^\circ$.
\item
The transverse magnetic anisotropy breaks the molecule's rotational symmetry around the easy axis $z$ (see also App.~\ref{sec:Magnetic_anisotropy}).
In consequence, one expects that such a symmetry breaking should manifest itself in different transport characteristics of the system occurring for various orientations of the magnetic field in the hard plane (i.e., the plane perpendicular to the easy axis).
From \Figs{Fig_5}(a)-(b) it is clear that the sole  position dependence in practice does not allow one to derive reliably either the transverse magnetic anisotropy constant $E_n$ or the angle $\phi$. For this purpose, also the \emph{amplitude} of the Coulomb peak has to be taken into consideration.
\item
The  presence of transverse magnetic anisotropy can be confirmed by observation of how the field dependence of the Coulomb peak amplitude changes when
 rotating the field orderly in the hard plane, or in other words by varying the angle $\phi$.
Specifically, one should notice then \emph{two} significantly different shapes of the amplitude showing up every 90$^\circ$, cf. red lines with others in the right panel of \Fig{Fig_5}. These two limiting cases represent the situation when the magnetic field lies either along the molecule's hard axis~$x$ ($\phi=0^\circ$ or $\phi=180^\circ$), \Fig{Fig_5}(c,e), or along the molecule's intermediate axis~$y$ ($\phi=90^\circ$ or $\phi=270^\circ$), \Fig{Fig_5}(d,f). Consequently, this enables one to determine the approximate value of the angle~$\phi$.
\item
The effect of transverse magnetic anisotropy on the Coulomb peak amplitude should be most pronounced for the magnetic field aligned along the molecule's hard axis $x$, see App.~\ref{sec:Magnetic_anisotropy} and Figs.~\ref{Fig_2}(c)-(d) of the main article. For a sufficiently high temperature $T$ [see \Fig{Fig_5}(f)-(g) and \Fig{Fig_16}] one observes then formation of local maxima, marked by red arrows in \Figs{Fig_5}(c,e), from whose position the value of the transverse magnetic anisotropy constant $E_n$ can be numerically estimated.
\end{enumerate}

Importantly, the method under discussion relies on a simultaneous fitting of \emph{position} (sensitive to $D_n$) and the \emph{amplitude} (sensitive both to $D_n$ and $E_n$) of the Coulomb peak. This strictly limits the freedom of the parameters' choice, basically leaving $E_n$ to be determined from the field value at which the maximum amplitude is acquired.
For instance, making the parameters $D_n$ smaller by 25\% than the one used above (given the fixed experimental temperature $T=1.8$ K), while assuming $E_n=0$,
may also produce a maximum, see green lines in \Figs{Fig_5}(c)-(f). However, not only does it result in  peak positions at completely wrong magnetic fields [cf. position of green and red arrows in \Fig{Fig_5}(e)], but also the amplitude shape remain unaltered upon changing the orientation of the field in the hard plane [cf. red and green lines between \Figs{Fig_5}(e) and (f)].
This restriction, combined with the sensitivity of the qualitative curve shape of the conductance to the parameters is advantageous for extracting the anisotropy parameters of SMMs \emph{in situ}.

\section{Conclusions}
In conclusion, we have proposed a new method of probing the transverse magnetic anisotropy of an individual SMM embedded in a three-terminal device.
It exploits the information contained in the spin states of the molecule through the analysis of
 the magnetic field evolution of the linear conductance amplitude $G_\text{max}$. We found that the evolution of $G_\text{max}$ in a magnetic field could only be reproduced when including a sufficient number of excited states.
Estimates for  the transverse anisotropy of the Fe$_4$ SMM yield $E \approx 0.17 D = 9.5$ $\mu$eV, a value of $E$ significantly larger than the observed bulk/monolayer values.
This is expected for a molecule
captured
 in the low symmetry environment of a transport junction.
Importantly, the technique does not rely on the small induced tunneling effects and hence works well at temperatures by far exceeding the tunnel splittings and even $E$ itself.
Our measurements find larger modulation of $G_\text{max}$ than calculated and the origin of this enhancement requires further study.
This method may facilitate the detection of \emph{in-situ} mechanical tuning~\cite{Grose_NatureMater.7/2008} or excitation~\cite{May_BeilsteinJ.Nanotechnol.2/2011,Burzuri_NanoLett.14/2014} of magnetic anisotropy of a single molecule.

\acknowledgments

This work was supported by NWO (VENI) and OCW, and by the EU FP7 project
618082 ACMOL and advanced ERC grant (Mols@Mols). M.M. acknowledges the financial support from the Alexander von Humboldt Foundation. K.P. was supported by U.S. National Science
 Foundations DMR-1206354.

\appendix
\section{\label{sec:Experiment}Materials and experimental method}
\subsection{\label{sec:Fe4_SMM}Details of the Fe$_{4}$ single-molecule magnet:}

\begin{figure}[t!!!]
    \includegraphics[width=0.98\columnwidth]{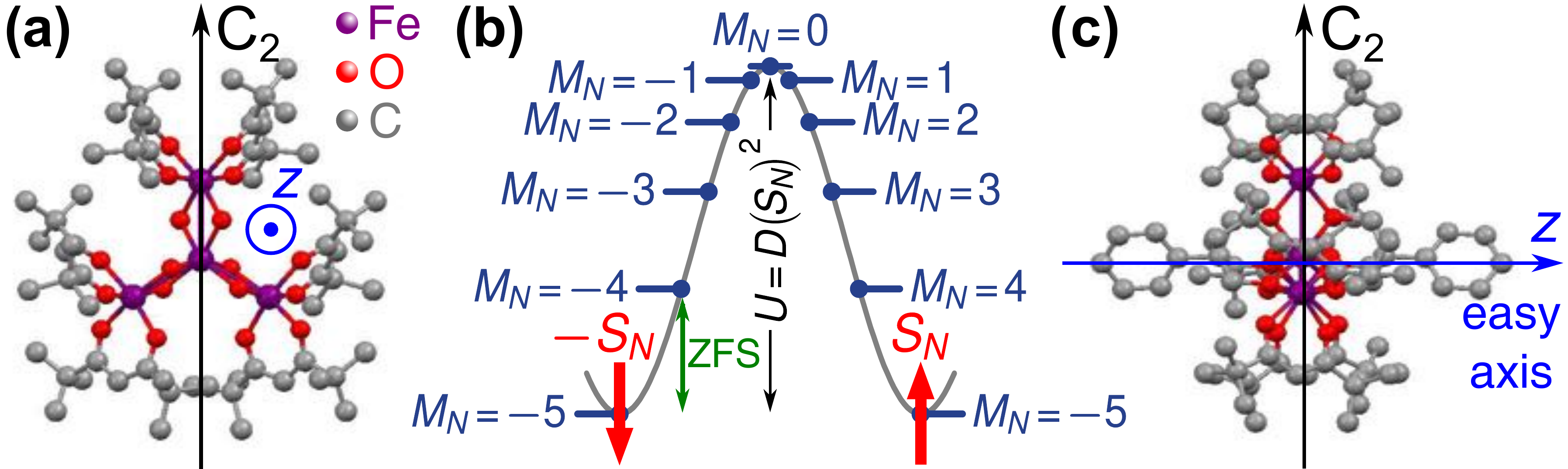}
    \caption{
    (color online)
    Details of the Fe$_{4}$ single-molecule magnet:
    (a) Sketch of the magnetic core of the Fe$_{4}$ SMM.
    (b) Ground-state spin multiplet ($S_N=5$) of the Fe$_{4}$ SMM in a neutral charge state $N$ -- for further explanation see App.~\ref{sec:Magnetic_anisotropy}.
    (c) Depiction of the Fe$_{4}$ SMM illustrating the orientation of the phenyl rings [omitted in (a)] that terminate the molecule. Note that both in (a) and (c) hydrogen atoms are disregarded for clarity.
    }
    \label{Fig_6}
\end{figure}

We used an Fe$_{4}$ SMM with formula [Fe$_{4}$(L)$_{2}$(dpm)$_{6}$]$\cdot$Et$_{2}$O where Hdpm is 2,2,6,6-tetramethyl-heptan-3,5-dione and H$_{3}$L is the tripodal ligand 2-hydroxymethyl-2-phenylpropane-1,3-diol, which carries a phenyl substituent.~\cite{Accorsi_J.Am.Chem.Soc.128/2006} In the bulk phase, the crystallographic symmetry is $C_{2}$.~\cite{Accorsi_J.Am.Chem.Soc.128/2006} The magnetic core of the Fe$_{4}$ SMM is made of 4 Fe$^{3+}$ ions (each with spin $s= 5/2$) as illustrated in Fig.~\ref{Fig_6}(a). The antiferromagnetic exchange interaction between the central and peripheral ions yields a large molecular spin $S_N= 5$ in the ground state. Magnetic anisotropy due to the interaction with the crystal field lifts the degeneracy of the spin multiplet into five doublets and one singlet that are distributed over an energy barrier as shown in Fig.~\ref{Fig_6}(b) -- for further discussion see App.~\ref{sec:Magnetic_anisotropy}. The height of the barrier, which hinders the spin reversal, is given by $U=D(S_N)^{2}$, where $D$ is the uniaxial magnetic anisotropy parameter. In the case of bulk Fe$_{4}$ the height is $U$=1.4 meV.~\cite{Accorsi_J.Am.Chem.Soc.128/2006} The `\emph{zero-field splitting}' (ZFS), defined as the energy difference between the two lowest-lying doublets ($M_N=\pm5$ and $M_N=\pm4$) is 0.5 meV. The low symmetry of the molecule induces a transverse magnetic anisotropy $E$ that, in bulk, is $E= 2.85$ $\mu$eV from EPR measurements.~\cite{Accorsi_J.Am.Chem.Soc.128/2006} Finally, we note that the molecule contains two axial tripodal ligands $L^{3-}$ which hold the core together and six peripheral dpm$^-$ ligands that create an hydrophobic envelope, see Fig.~\ref{Fig_6}(c).

\subsection{\label{sec:Three-terminal_junctions}Details on the fabrication methods of the three-terminal junctions}

\begin{figure}[t]
    \includegraphics[width=0.925\columnwidth]{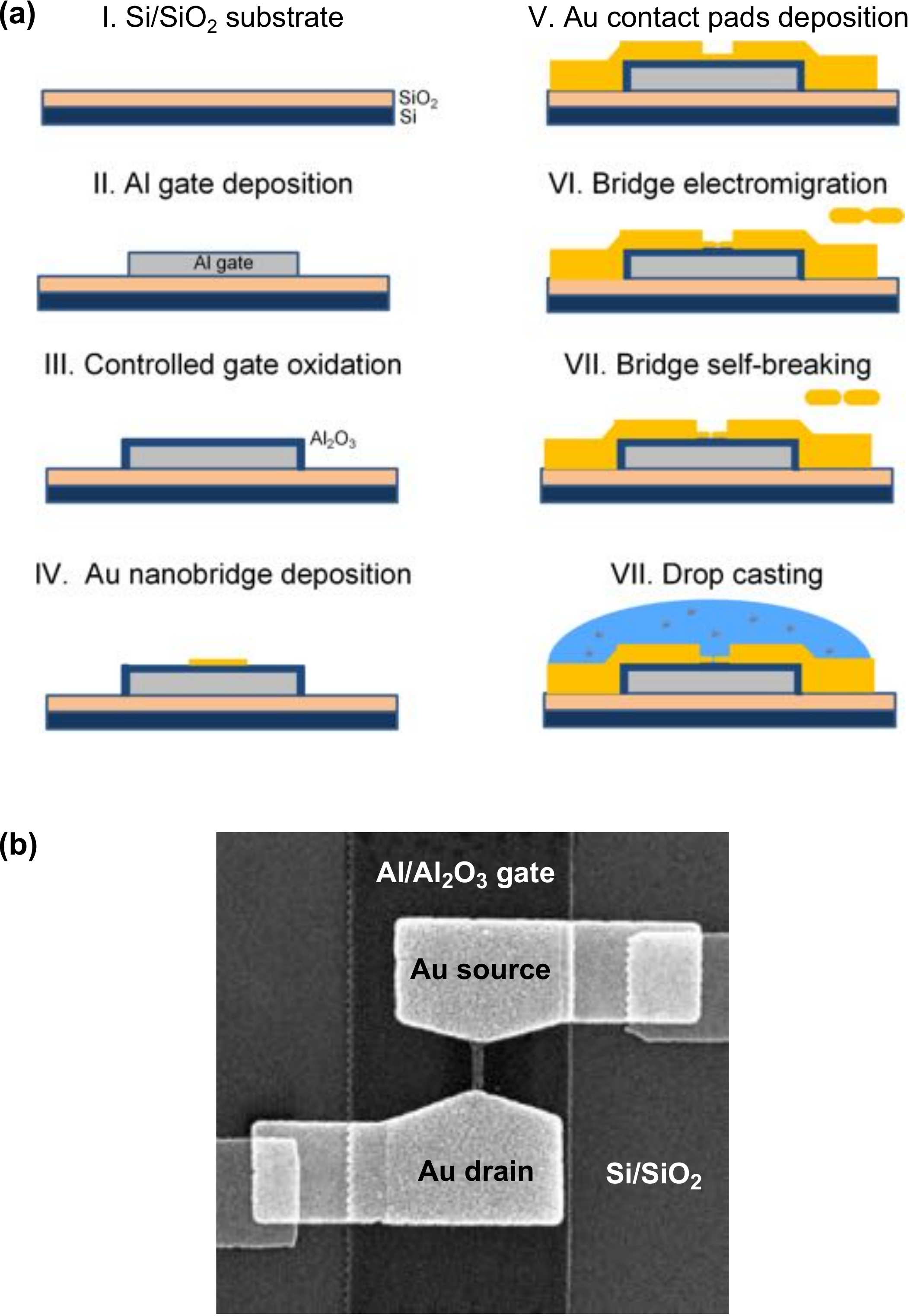}
    \caption{
    (color online)
    Three-terminal-junction fabrication:
    (a) Schematics of the three-terminal-device fabrication process.
    (b) Scanning electron microscope (SEM) image of a real three-terminal device before electromigration.
    }
    \label{Fig_7}
\end{figure}

The three-terminal junctions are fabricated on a silicon substrate covered by 280 nm of SiO$_2$. The schematics of the fabrication process is described in Fig.~\ref{Fig_7}(a). The gate electrode is fabricated by e-beam lithography and subsequent e-beam deposition of Al. In the next step, the  oxidation of the gate in a controlled oxygen atmosphere produces a dielectric coating layer of 2-3 nm of Al$_2$O$_3$.
The source and drain electrodes are fabricated by self-breaking, controlled electromigration of a Au nanobridge deposited by e-beam lithography on top of the oxidized gate. The self-breaking technique prevents the formation of gold nano-grains in the junction that could mimic the behavior of a molecule. Figure~\ref{Fig_7}(b) shows a scanning electron microscope image of a device before electromigration.

The molecules are deposited onto the chip by drop casting a $10^{-4}$ M solution in toluene into a liquid cell containing the chip with the junctions. The electromigration of the bridge and subsequent self-breaking are carried out in solution to maximize the yield of junctions with a molecule.

\subsection{\label{sec:Gate-voltage_position}Details on the gate-voltage `position' spectroscopy}

The molecule-electrode coupling $\Gamma$ is estimated from the broadening of the Coulomb edge at low bias.
In particular, the full-width at half-maximum (FWHM) of the Coulomb peak is used for this purpose. We find 1.6 meV, 2.0 meV and 1.4 meV for samples A, B and C respectively.
Note, however, that these values are an upper limit for $\Gamma$ since we cannot resolve the presence of additional components for the broadening such as
thermal energy
 or the contribution of other molecular levels very close in energy.

\begin{figure}[t]
    \includegraphics[width=0.99\columnwidth]{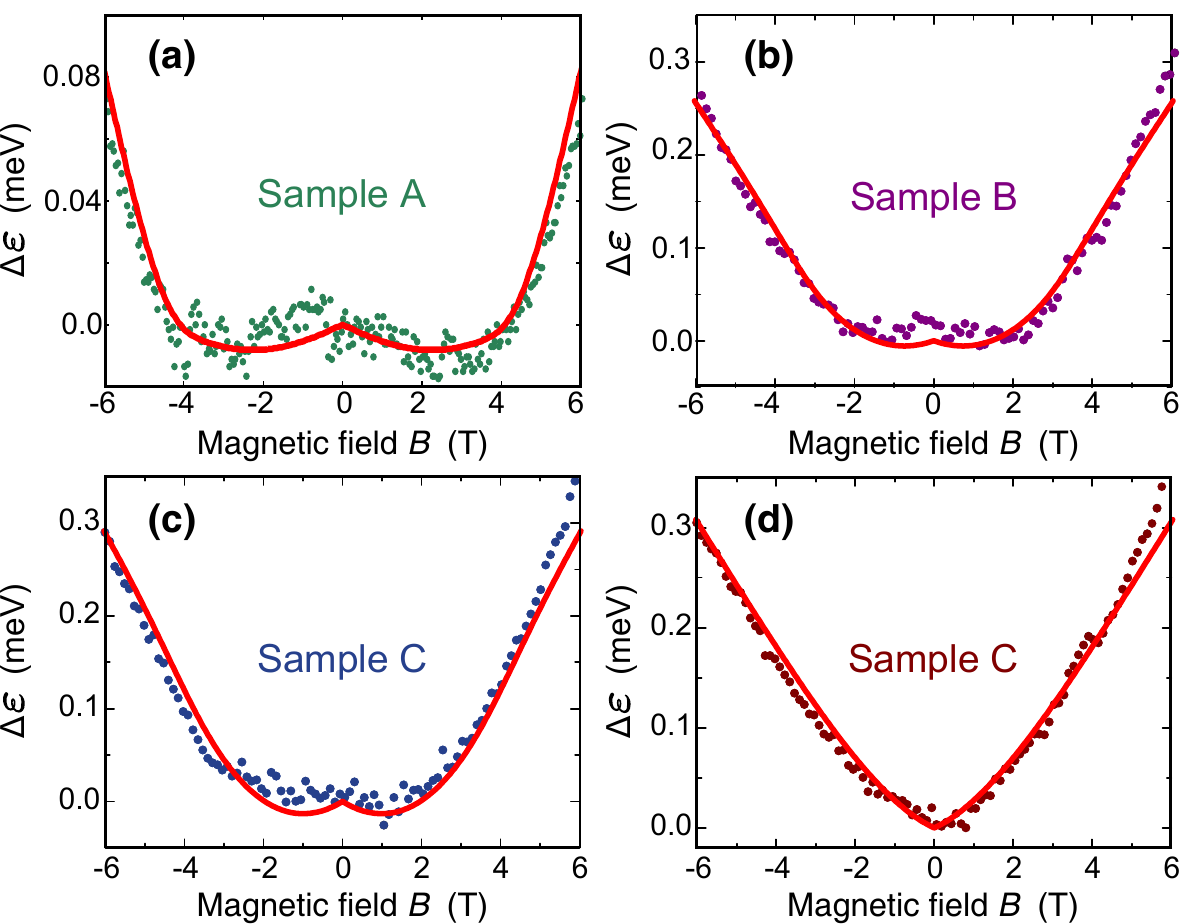}
    \caption{
    (color online)
    Coulomb peak position gate-voltage spectroscopy:
     The shift of the Coulomb peak position due to magnetic field for samples A, B and C. The solid lines are fits to $\varepsilon^{0}_{N+1}-\varepsilon^{0}_{N}$, calculated from the giant-spin Hamiltonian,
     Eqs.~(\ref{eq:H_SMM})-(\ref{eq:H_N}).
     From the fit we get the following values --
    for sample A in (a):
    $D_{N+1}=61$ $\mu$eV , $\theta_{N}= 87^{\circ}$ and $\theta_{N+1}= 86^{\circ}$;
    for sample B in (b): $D_{N+1}=65$ $\mu$eV, $\theta_{N}=86^{\circ}$ and $\theta_{N+1}=84^{\circ}$;
     for sample C: in (c) $\theta_{N}=87^{\circ}$ and $\theta_{N+1}=85^{\circ}$, whereas in (d) $\theta_{N}=63^{\circ}$ and $\theta_{N+1}=62^{\circ}$,
     with $D_{N+1}=68~\mu$eV in both cases.
     We note that the evolution of the Coulomb peak position in magnetic field, and not  $G_\text{max}$, for samples A and C was previously analyzed in~Ref.~[\onlinecite{Burzuri_Phys.Rev.Lett.109/2012}].
     Also note that  in the fitting for sample A  we included $E/D=0.2$ and $\phi=0^{\circ}$ obtained in
     Fig.~\ref{Fig_2}.
     }
    \label{Fig_8}
\end{figure}

\begin{figure*}[t!!!]
    \includegraphics[width=0.8\textwidth]{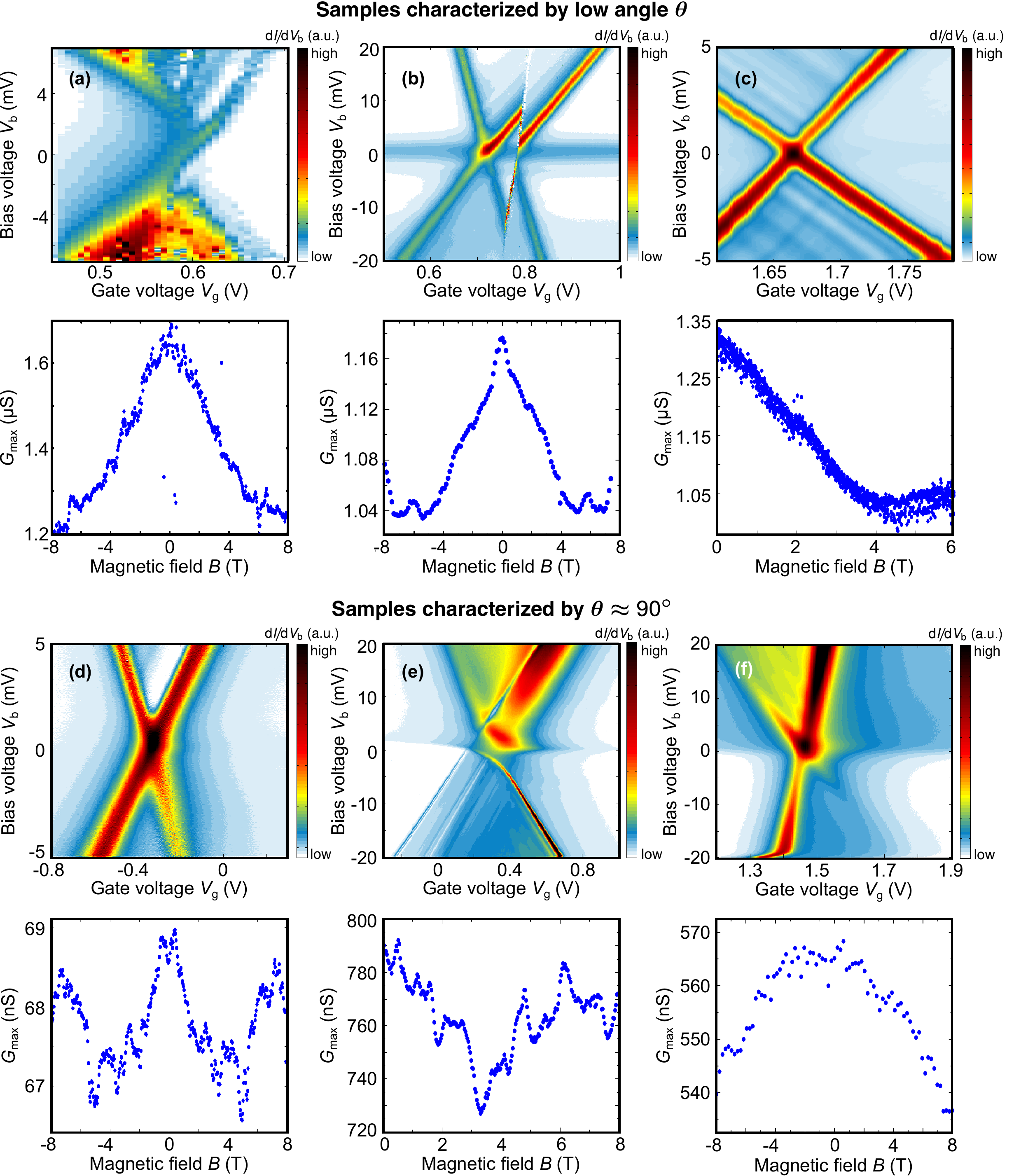}
    \caption{
    (color online)
    Statistics:
    differential-conductance maps, $\text{d}I/\text{d}V_\text{b}$, shown as a function of gate $V_\text{g}$ and bias $V_\text{b}$ voltages together with corresponding dependencies of the Coulomb peak amplitude $G_\text{max}$ on magnetic field $B$ for six different Fe$_4$ molecular junctions.
    \emph{Top panel} [(a)-(c)]: Junctions for which gate-voltage spectroscopy fits of the Coulomb peak position (not shown) indicate $\theta<60^\circ$.
    \emph{Bottom panel} [(d)-(f)]:
    Junctions where $\theta\approx90^{\circ}$ is found. The shape of the field modulation of $G_\text{max}$ implies that for (d) and (e) the field is close to the intermediate axis ($\phi\approx90^\circ$), whereas for (f) it is most likely in an intermediate $\phi$ angle in the hard plane.
    }
    \label{Fig_9}
\end{figure*}

Figure~\ref{Fig_8} shows the Coulomb peak (CP) position in gate voltage $V_\text{g}$ as a function of the magnetic field for the samples A, B and C described in the main text. The values of  $V_\text{g}$ are multiplied by the gate coupling $\beta$ to obtain energy units ($\Delta \varepsilon$) and subsequently re-scaled to make $\Delta \varepsilon=0$ for $B=0$. The non-linearity of the field dependence is a clear signature of the magnetic anisotropy as described in the text (see also Ref.~[\onlinecite{Burzuri_Phys.Rev.Lett.109/2012}]).
Moreover, the low-field `flatness' of $\Delta \varepsilon$ observed in Figs.~\ref{Fig_8}(a)-(c) is indicative of a high value of $\theta$ in contrast with Fig.~\ref{Fig_8}(d). The solid lines in Fig.~\ref{Fig_8} are a fit of the data to $\Delta \varepsilon=\varepsilon^{0}_{N+1}-\varepsilon^{0}_{N}$ as defined by the giant-spin Hamiltonian,   Eqs.~(\ref{eq:H_SMM})-(\ref{eq:H_N}),
and also discussed in detail in App.~\ref{sec:Model}.
The CP position is mainly insensitive to $E$  (see also Supporting Information of Ref.~[\onlinecite{Burzuri_Phys.Rev.Lett.109/2012}]), and therefore we can independently extract the parameters $D$ and $\theta$ related to the uniaxial anisotropy. Note that we fix the value of $D_N$ (neutral state) to the bulk value $D_{N}= 56$ $\mu$eV and thus the free parameters are $D_{N+1}$, $\theta_{N}$ and $\theta_{N+1}$. See the caption of Fig.~\ref{Fig_8} for the fitting values of these parameters.

\subsection{\label{sec:Statistics} Statistics and effect of the magnetic field polarity}

We measured around 200 electromigrated junctions from which 17 showed molecular signatures. A total of 9 molecular junctions displayed a clear Coulomb peak suitable for further analysis by means of the gate-voltage spectroscopy method, from which the junctions were proven to exhibit magnetic anisotropy. Importantly, all these junctions displayed a modulation of the peak amplitude $G_{\text{max}}$ as a function of the magnetic field. A total of 6 of these samples could be rotated or were close to $\theta=90^{\circ}$. From those, one sample was close to $\phi=0^{\circ}$ (hard axis), and it is referred to as sample A.
Figure~\ref{Fig_9} shows the differential-conductance maps, $\text{d}I/\text{d}V_\text{b}$,  and corresponding  magnetic field evolutions of $G_\text{max}$ for different Fe$_4$ molecular junctions, that is other than samples A, B and C discussed in the main text.
The top panel [(a)-(c)] of Fig.~\ref{Fig_9} presents samples for which the gate spectroscopy yields low values of $\theta$.
Worthy of note is that for $|B|<4$ T a decrease of $G_\text{max}$ is observed with increasing $|B|$.
On the other hand, the bottom panel [(d)-(f)] of  Fig.~\ref{Fig_9} shows examples where $\theta\approx90^{\circ}$ (i.e., close the the hard plane). The shape of $G_\text{max}$  for (d) and (e) indicates that the magnetic field is close to the intermediate axis ($\phi\approx90^\circ$), which follows from the analysis carried out in the main text.
For the last sample, Fig.~\ref{Fig_9}(f), the field is most likely at an intermediate angle $\phi$ in the hard plane.

\begin{figure}[t!!!]
    \includegraphics[width=0.99\columnwidth]{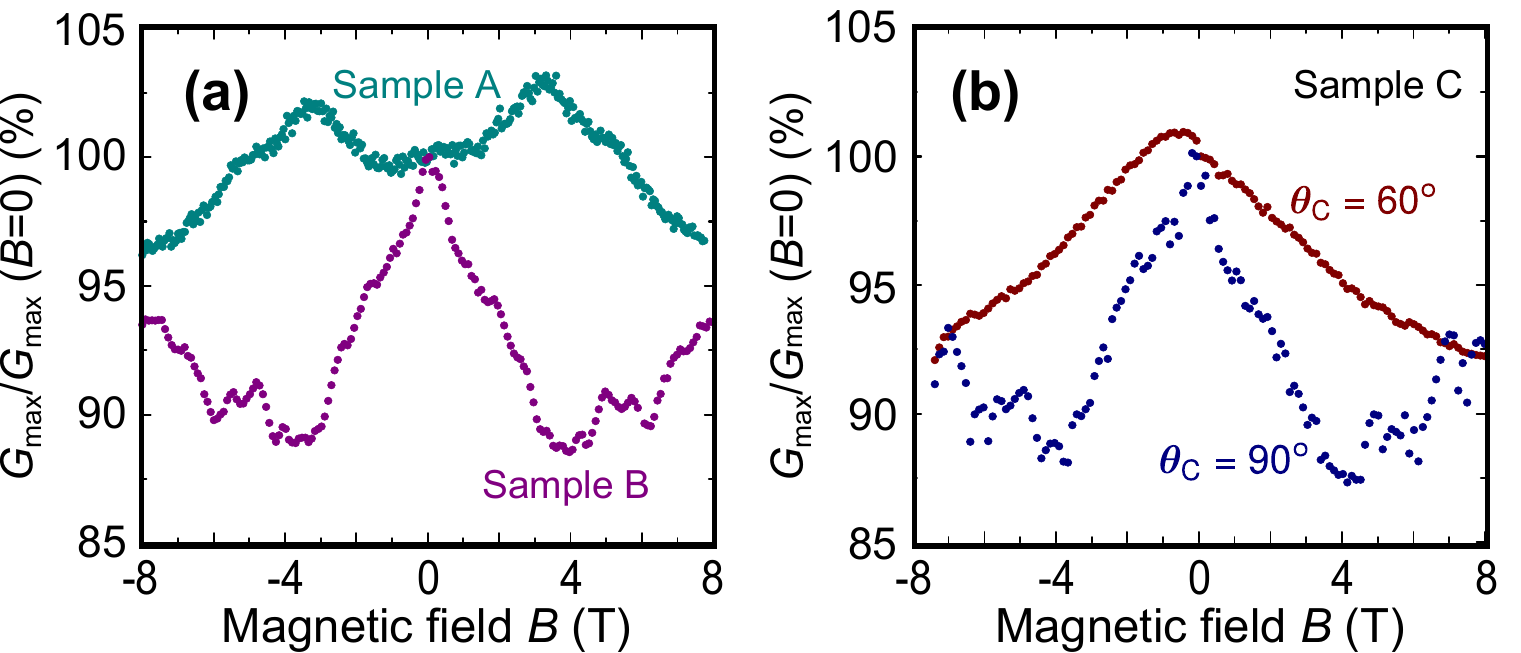}
    \caption{
    (color online)
    The effect of the reversed magnetic field polarity on $G_\text{max}$:
    Dependence of the scaled Coulomb peak height $G_\text{max}/G_\text{max}(B=0)$ on magnetic field $B$ for the samples discussed in the main text, cf.~Fig.~\ref{Fig_2}(a)-(b),  showing that the curves are symmetric upon reversal of the field polarity.
    }
    \label{Fig_10}
\end{figure}

In order to discard the influence of universal conductance fluctuations induced by the magnetic field in the measurements, in Fig.~\ref{Fig_10} we plot $G_{\text{max}}$ as a function of $B$ for the samples shown in Figs.~\ref{Fig_2}(a)-(b) for both positive and negative polarities of magnetic field. We note that the main features, like the minima or maxima around~4~T, are reproducible under inversion of the field polarity. Universal conductance fluctuations are not expected to be symmetric by changing the $B$ polarity. Some additional symmetric structure appears also in the measurements. The analysis of this smaller contribution is interesting but beyond the scope of this work.

\begin{figure}[t!!!]
    \includegraphics[width=0.99\columnwidth]{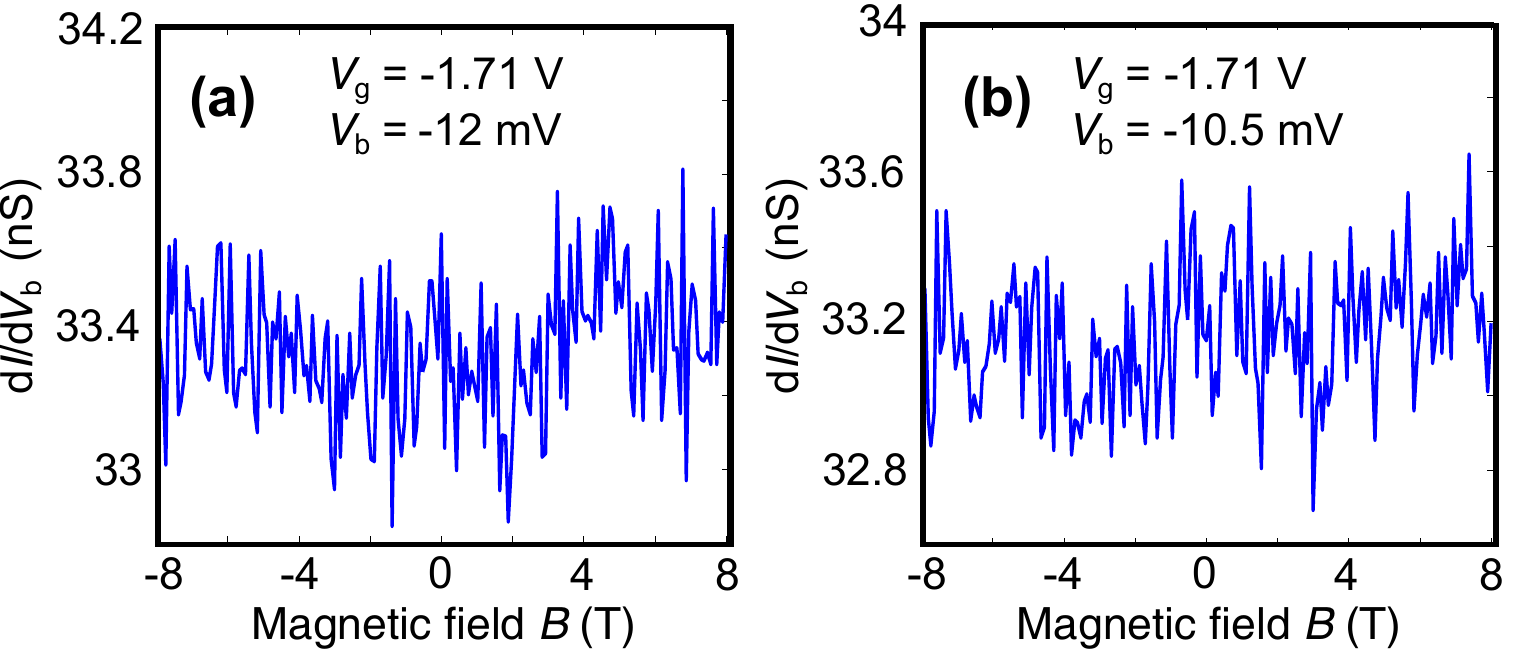}
    \caption{
    (color online)
    Co-tunneling background:
    differential conductance, $\text{d}I/\text{d}V_\text{b}$, measured as a function of magnetic field $B$ at two different points: (a) $V_{\text{g}}=-1.71$ V and $V_\text{b}=-12$ mV and (b) $V_{\text{g}}=-1.71$ V and $V_\text{b}=-10.5$ mV, which correspond to the co-tunneling background in the left-hand charge state of Sample~A, cf. Fig.~\ref{Fig_1}(c).
    }
    \label{Fig_11}
\end{figure}

If present, conductance fluctuations would equally appear in the zero-bias and the higher bias conductance. Therefore, in order to rule out their presence, we have analyzed the magneto-resistance at higher biases and different gate voltages. Figure~\ref{Fig_11} shows differential conductance,  $\text{d}I/\text{d}V_\text{b}$, as a function of $B$ measured at two different and fixed gate $V_\text{g}$ and bias $V_\text{b}$ voltages in the Coulomb blockade in Sample A. We observe an almost flat response of $\text{d}I/\text{d}V_\text{b}$ with peak-to-peak variation of the order of 0.1 nS. This magnitude is not comparable to the modulations we attribute to the presence of the transverse anisotropy.  Moreover, note that these two spectra are not symmetric by reversing the magnetic field polarity. Thus, we conclude that the universal conductance fluctuations are not significant in our measurements.

\section{\label{sec:Theory}Theoretical modelling}

\begin{figure*}[t]
   \includegraphics[scale=0.85]{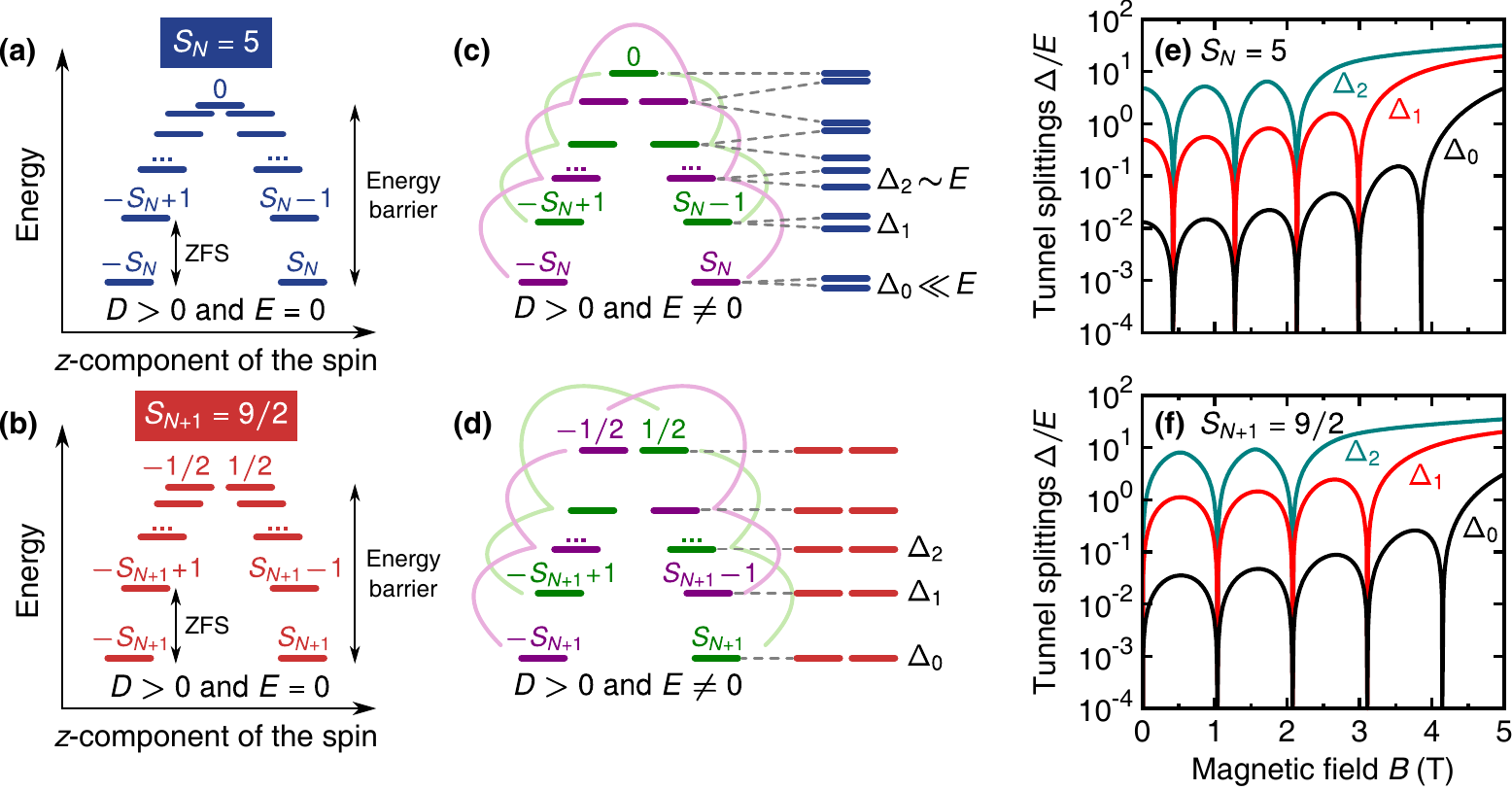}
    \caption{
    (color online)
    Effect of magnetic anisotropy on the energy spectrum of SMM:
    \emph{Top}/\emph{Bottom panel} [(a,c,e)/(b,d,f)] illustrates the case of a \emph{integer}/\emph{half-integer} value of a molecular spin.
    In particular, we use the values of spin known for a Fe$_4$ SMM, $S_N=5$ for a neutral molecule and $S_{N+1}=9/2$ for a charged one.\cite{Zyazin_NanoLett.10/2010}
    (a)-(b)
    In the presence of exclusively \emph{uniaxial} magnetic anisotropy $D>0$ (and without magnetic field, $B=0$)  an energy barrier protecting the molecule's spin against reversal between two opposite, energetically degenerate, orientations arises. The excitation between the ground state doublet and the first excited doublet is then commonly referred to as the \emph{`zero-field splitting'} (ZFS).
    (c)-(d)
    If additionally the \emph{transverse} component of magnetic anisotropy occurs, it allows for mixing of pure $S_z$-states. Each new eigenstate is then formed from $S_z$-states belonging to one of two uncoupled, time-reversed sets, as schematically marked by two different colors. As follows from the Kramers theorem, for $S_N=5$ the transverse magnetic anisotropy introduces tunnel-splittings $\Delta$, whereas for $S_{N+1}=9/2$ all states remain doubly degenerate.
    (e)-(f)
    A characteristic feature of such anisotropic, large spins is that when an external magnetic field $B$ is applied along the system's hard axis, one observes periodic changes of the tunnel-splittings.\cite{Wernsdorfer_Science284/1999,Gatteschi_book} Other parameters assumed in the calculations: $D_N=56$ $\mu$eV, $D_{N+1}=68$ $\mu$eV, and $E_{N}/D_N=E_{N+1}/D_{N+1}=0.3$.
  }
  \label{Fig_12}
\end{figure*}

\subsection{\label{sec:Model}Charge-dependent, giant-spin-based model of an single-molecule magnet}
The central element of the theoretical description of the gate-spectroscopy technique is a proper choice of the model capturing essential features of an SMM.
As introduced in the main article, the molecule is represented by a model based on two giant-spin Hamiltonians.~\cite{Kahn_book,Boca_book,Gatteschi_book}  This allows us to  take into account the fact that by tuning a gate voltage $V_\text{g}$ the molecule can be switched between two different charge states,~\cite{Zyazin_NanoLett.10/2010} referred to as a \emph{neutral} ($N$) and \emph{charged} ($N+1$) one. In general, each of this states can be characterized not only by different values of molecular ground-state spin ($S_N$ and $S_{N+1}$), but also uniaxial ($D_N$ and $D_{N+1}$) and transverse ($E_N$ and $E_{N+1}$) magnetic anisotropy constants. Using the  spin raising/lowering operators $\hat{S}_n^\pm$, the Hamiltonian of an SMM in the  charge  state $n$ and subject to an arbitrarily oriented external magnetic field $\vec{B}$ takes the form
given by Eqs.~(\ref{eq:H_SMM})-(\ref{eq:H_N}) and the Zeeman term explicitly given by
    \begin{multline}\label{eq:H_Z_n}
    \hat{\mathcal{H}}_{n}^\text{Z}
    =
    g\mu_\text{B}B
    \Big[
    \frac{1}{2}\hat{S}_n^+\sin\theta\,\text{e}^{-i\phi}
    \\
    +
    \frac{1}{2}\hat{S}_n^-\sin\theta\,\text{e}^{i\phi}
    +
    \hat{S}_n^z\cos\theta
    \Big]
    ,
    \end{multline}
with the angles $\theta$ and $\phi$ defined as illustrated in Figure~1b. Noteworthily, by keeping the same value of $\theta$ and $\phi$ for both charge states, we implicitly assume that the orientation of the molecule's principle axes set by magnetic anisotropy is not affected by charging. This assumption not necessarily holds for real systems as shown in Refs.~[\onlinecite{Burzuri_Phys.Rev.Lett.109/2012}] and [\onlinecite{Zyazin_NanoLett.10/2010}]. However, since the tilting, if observed, usually does not exceed few degrees, we do not include such an effect into the present considerations.

\subsection{\label{sec:Magnetic_anisotropy}How does magnetic anisotropy affect the energy spectrum of a large spin?}

Before we analyze how electronic transport probes the transverse magnetic anisotropy of a molecule, it may be instructive first to  discuss the consequences of the transverse magnetic anisotropy and external magnetic field for the SMM's energy spectrum.

To begin with, as long as the transverse magnetic anisotropy is vanishingly small the system can be described simply by the first term of the Hamiltonian (\ref{eq:H_N}). As a result, the eigenvalues $M_n$ of the spin operator $\hat{S}_n^z$ become good quantum numbers for labelling the eigenstates of $\hat{\mathcal{H}}_{\text{SMM},n}=-D_n\big(\hat{S}_n^z\big)^{\!2}$, that is $\hat{\mathcal{H}}_{\text{SMM},n}\ket{M_n}=-D_nM_n^2\ket{M_n}$. For $D_n>0$ the energy spectrum of an SMM takes the form of an inverted parabola with an energy barrier of height $\sim D_nS_n^2$ for spin reversal, which basically corresponds to the indirect transition between the ground states $\ket{-S_n}$ and $\ket{S_n}$ by climbing the barrier \emph{via} the intermediate states $\ket{M_n}$ (for $M_n=-S_n+1,\ldots,S_n-1$), see Figs.~\ref{Fig_12}(a)-(b). Importantly, the excitation energy between the ground state $\ket{\pm S_n}$ and the first excited state $\ket{\pm S_n \mp1}$, the so-called \emph{`zero-field splitting'} $\text{ZFS}=(2S_n-1)D_n$, sets the threshold energy scale for the reversal process to take place.
Note that transition energies between neighboring excited states $\ket{M_n}$ and $\ket{M_n^\prime}$ with $\big|M_n-M_n^\prime\big|=1$ are characterized by energies $(2M_n-1)D_n$ (for $0<M_n<S_n$) that are smaller than the ZFS, and these  states remain generally unpopulated until  the ground-to-first excited state transition becomes energetically permitted. This bottleneck behavior manifest then in electronic transport through an SMM, where it can be observed as a step-like feature in the conductance only at bias voltages $V_\text{b}=\pm \text{ZFS}/|e|$.~\cite{Zyazin_NanoLett.10/2010,Misiorny_Phys.Rev.Lett.111/2013}

The relatively simple picture presented above is not valid, however, if the transverse magnetic anisotropy (or an external magnetic field perpendicular to the molecule's easy axis) is significant. When $E\neq0$, the second term of the Hamiltonian~(\ref{eq:H_N}) breaks the system's rotational symmetry around the easy axis $z$, so that $M_n$ is no longer a good quantum number. In fact, each of the $2S_n+1$ eigenstates of $\hat{\mathcal{H}}_{n}=-D_n\big(\hat{S}_n^z\big)^{\!2} +(E_n/2)\big[\big(\hat{S}_n^+\big)^{\!2}+\big(\hat{S}_n^- \big)^{\!2}\big]$ is now a linear combination of the eigenstates $\ket{M_n}$, which, in turn, underlies the origin of the quantum tunneling of magnetization.~\cite{Gatteschi_Angew.Chem.Int.Ed.42/2003} In particular, each of these eigenstates is formed from states $\ket{M_n}$ belonging to one of two uncoupled, time-reversed sets, as shown in Figs.~\ref{Fig_12}(c)-(d). For an \emph{integer} spin $S_n$, the transverse magnetic anisotropy  leads to splitting of energy levels, usually referred to as \emph{`tunnel-splittings'}, Fig.~\ref{Fig_12}(c), whereas for a \emph{half-integer} spin $S_n$ (in the absence of magnetic field) according to the Kramers theorem each energy level is doubly degenerate, Fig.~\ref{Fig_12}(d). Interestingly, if one applies an external magnetic field in the direction perpendicular to the system's easy axis $z$, periodic changes of these tunnel-splittings can be observed if the field is oriented along or close the hard axis $x$, Figs.~\ref{Fig_12}(e)-(f), and they disappear as the field gets rotated towards the direction of the medium axis~$y$.\cite{Gatteschi_book,Wernsdorfer_Science284/1999,Gatteschi_Angew.Chem.Int.Ed.42/2003}

\subsection{\label{sec:Transport}Transport in the SET regime}

For a weak tunnel-coupling between an SMM and electrodes, transport in the single electron tunneling (SET) regime can be considered in the leading-order perturbative approach (Fermi golden rule combined with a master equation).\cite{Romeike_Phys.Rev.Lett.96/2006_196805,Timm_Phys.Rev.B73/2006,Misiorny_Phys.Rev.B79/2009}

We describe metallic, \emph{nonmagnetic} electrodes [$q=(L)\text{eft},(R)\text{ight}$] as reservoirs of noninteracting electrons, whose tunneling processes to/from a molecule are modelled by the following Hamiltonian
    \begin{align}
    \hat{\mathcal{H}}_\text{tun}
    =\ &
    \sum_{qkl\sigma}
     t_{l}^q
     \hat{d}_{l\sigma}^\dagger \hat{a}_{k\sigma}^q
     +
     \text{H.c.}
     \nonumber\\
    =\ &
    \sum_{qk\sigma}
    \sum_{a_Nb_{N+1}}
    T_{a_{N+1}b_N}^{\sigma q} \ket{a_{N+1}}\bra{b_N}\hat{a}_{k\sigma}^q
    +
    \text{H.c.}
    \end{align}
with
    \begin{equation}
   T_{a_{N+1}b_N}^{\sigma q}
   =
   \sum_l t_{l}^q \bra{a_{N+1}}\hat{d}_{l\sigma}^\dagger\ket{b_{N}},
    \end{equation}
where $t_l^q$ is the tunneling matrix element, $\hat{d}_{l\sigma}^\dagger$ represents creation of an electron with spin $\sigma$ in the molecular orbital~$l$, and $\hat{a}_{k\sigma}^q$ denotes the annihilation operator for the $q$th electrode with $k$ standing for an orbital quantum number. Note that the molecular state has been expanded in the basis of eigenvectors $\ket{a_{N+1}}$ and $\ket{b_N}$ of $\hat{\mathcal{H}}_{\text{SMM}}=\sum_{n=N,N+1}\hat{\mathcal{H}}_{\text{SMM},n}$.
Next, we express the molecular eigenstates $\ket{a_N}$ and $\ket{b_{N+1}}$ with respect to the basis of angular momentum (spin) eigenstates. In principle, an arbitrary molecular state can be decomposed as $\ket{\chi_n}=\sum_{S_nM_n}\chi_{S_nM_n}\ket{S_nM_n}$. As a result, one obtains
   \begin{multline}
   T_{a_{N+1}b_N}^{\sigma q}
   =
    \sum_l
    \sum_{S_{N+1}M_{N+1}}
    \sum_{S_N M_N}
    t_{l}^q
    a_{S_{N+1}M_{N+1}}^\ast
    b_{S_N M_N}^{\phantom{\ast}}
    \\
    \times
    \bra{S_{N+1}M_{N+1}}d_{l\sigma}^\dagger\ket{S_N M_N}.
   \end{multline}
The key problem one encounters when analyzing the above equation is that the operator $\hat{d}_{l\sigma}^\dagger$ involves two degrees of freedom, namely, the orbital one ($l$) and the spin one ($\sigma$). Consequently, it may seem that in the next step we need to
calculate $\bra{S_{N+1}M_{N+1}}\hat{d}_{l\sigma}^\dagger\ket{S_N M_N}$ explicitly.  This complication, however, can be avoided by making use of the the Wigner-Eckart theorem,~\cite{Messiah_book} which basically allows for finding matrix elements of an operator with respect to angular momentum eigenstates,
    \begin{multline}
    \hspace*{-10pt}
    \bra{S_{N+1}M_{N+1}}\hat{d}_{l\sigma}^\dagger\ket{S_N M_N}
    =
    \langle S_N,M_N;\tfrac{1}{2},\sigma|S_{N+1},M_{N+1} \rangle
    \\
    \times
    \bra{S_{N+1}}|\hat{d}_l^\dagger|\ket{S_N}.
    \end{multline}
The first factor of the RHS is a Clebsch-Gordan coefficient for adding spins $S_N$ and $1/2$ to get $S_{N+1}$. This depends only on how the system is oriented with respect ot the $z$ axis. On the other hand, the second factor, the so-called \emph{reduced matrix element}, remains independent of the spatial orientation, as it does not contain the magnetic quantum numbers $M_N$, $M_{N+1}$ or $\sigma$. Thus, we get
    \begin{align}
    T_{a_{N+1}b_N}^{\sigma q}
    =\ &
     \sum_{S_NS_{N+1}}
    \mathcal{T}_{a_{N+1}b_N}^{\sigma}
    \mathbb{T}_{S_{N+1}S_N}^q
    ,
    \end{align}
with
    \begin{multline}
    \mathcal{T}_{a_{N+1}b_N}^{\sigma}
    =
     \sum_{M_N M_{N+1}}
     a_{S_{N+1}M_{N+1}}^\ast
     b_{S_N M_N}^{\phantom{\ast}}
     \\
     \times
    \langle S_N,M_N;\tfrac{1}{2},\sigma|S_{N+1},M_{N+1} \rangle
    ,
    \end{multline}
and the term
    $
    \mathbb{T}_{S_{N+1}S_N}^q = \sum_l t_l^q \bra{S_{N+1}}| \hat{d}_l^\dagger|\ket{S_N}
    $
regarded in calculations as a \emph{single} free parameter to be adjusted for each electrode. Specifically, assuming a symmetric coupling between the molecule and two identical electrodes ($t_l^L=t_l^R$), the tunnel coupling takes the from $\Gamma_L=\Gamma_R=\Gamma/2$, where $\Gamma=2\pi\rho\big|\mathbb{T}_{S_{N+1}S_N}\big|^2$ and  $\rho$ denotes the constant, spin-independent density of states in electrodes.

The stationary current $I$ flowing through a molecule is calculated as $I=(I_L-I_R)/2$, where $I_q$ (for $q=L,R$) stands for the current flowing from the $q$th electrode to the molecule,
    \begin{multline}
    I_q=
    \frac{e\Gamma}{2\hbar}
    \sum_{nn^\prime}
    \sum_{a_{n^{\phantom{\prime}}}\!\!b_{n^\prime}}
    (n^\prime-n)
    f_q(\Delta \varepsilon_{b_{n^\prime}\!,a_{n^{\phantom{\prime}}}\!\!})
    \\
    \times
    \sum_{\sigma\in q}
    \big|\mathcal{T}_{b_{n^\prime}\!,a_{n^{\phantom{\prime}}}\!\!\!}^{\sigma}\big|^2
    \mathcal{P}_{a_{n^{\phantom{\prime}}}}
    .
    \end{multline}
where $\Delta \varepsilon_{b,a}=\varepsilon_b- \varepsilon_a$, and $f_q(\omega)=(1+\exp[(\omega-\mu_q)/(k_\textrm{B}T)])^{-1}$ is the Fermi-Dirac function of the $q$th electrode, with $T$ and    $\mu_{L(R)}=\mu_0\pm eV_\text{b}/2$  standing for  temperature and  the relevant electrochemical potential, respectively. The probabilities $\mathcal{P}_{a_n}$ of finding an SMM in a specific state~$\ket{a_n}$ are then derived from a stationary master equation.~\cite{Romeike_Phys.Rev.Lett.96/2006_196805} Finally, since SMMs are typically characterized by long spin coherence and spin relaxation times as a result of a weak spin-orbit and hyperfine coupling to the environment,\cite{Ardavan_Phys.Rev.Lett.98/2007,Bertaina_Nature453/2008,Bogani_NatureMater.7/2008}
 we neglect relaxation of the spin states due to processes other than due to the electron tunneling.

 In Fig.~\ref{Fig_3}(d), and also in \Figs{Fig_13}--\ref{Fig_16}, we present the current $I_r=(I_L^r-I_R^r)/2$ which includes first $r$ lowest-in-energy states in the spin multiplet of each charge state. We use this to show that many excited states in both charge state have to be taken into account in order to describe current correctly.
We define $I_q^r$ in the following way
    \begin{multline}
    I_q^r=
    \frac{e\Gamma}{2\hbar}
    \sum_{nn^\prime}
    \sum_{b_{n^\prime}}
    \sideset{}{'}\sum\limits_{a_{n^{\phantom{\prime}}}\!\!}^r
    (n^\prime-n)
    f_q(\Delta \varepsilon_{b_{n^\prime}\!,a_{n^{\phantom{\prime}}}\!\!})
    \\
    \times
    \sum_{\sigma\in q}
    \big|\mathcal{T}_{b_{n^\prime}\!,a_{n^{\phantom{\prime}}}\!\!\!}^{\sigma}\big|^2
    \mathcal{P}_{a_{n^{\phantom{\prime}}}}
    ,
    \end{multline}
with \scalebox{0.85}{$\sideset{}{'}\sum\limits_{a_{n^{\phantom{\prime}}}\!\!}^r$} denoting summation over states $\ket{a_n}$ in the charge state $n$ that is limited only to first $r$ states of lowest energy.

\subsection{\label{sec:Signatures_E}Signatures of the transverse anisotropy parameter $E$ without the Berry phase oscillations}

In Figs.~\ref{Fig_2}(a) and \ref{Fig_3} we discuss the initial increase of the current with magnetic field followed by a decrease.
The key insight of our calculations using the method described in the previous section (App.~\ref{sec:Transport}) is that  the mechanism for this effect is significantly enhanced and modified for $E\neq0$ giving rise to the characteristic $G_\text{max}$ curves shown in Fig.~\ref{Fig_2}.
Since this is at the basis of our scheme of detection, it deserves a further comment. In particular, the relation to the Berry phase oscillations which underlay most of the previously used techniques for determining the parameter~$E$.

\begin{figure}[p]
\vspace*{75pt}
   \includegraphics[width=0.85\textwidth]{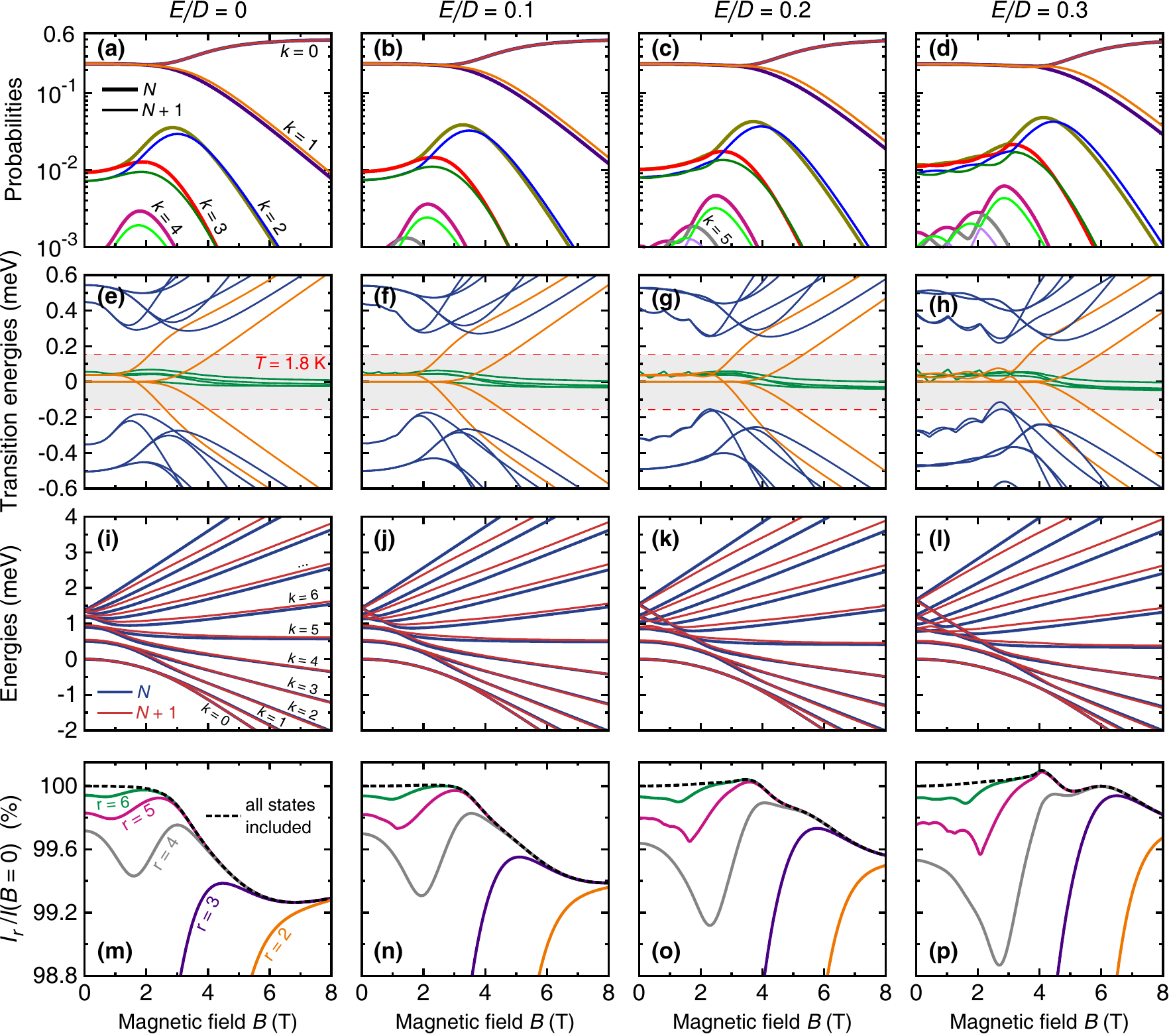}
    \caption{
    (color online)
    Signatures of the transverse magnetic anisotropy in electronic transport (magnetic field along the \emph{hard} axis, $\theta=90^\circ$ and $\phi=0^\circ$):
    Analogous to Figs.~\ref{Fig_3}(b)-(d) with each column corresponding now to a different value of $E/D$:
    (a)-(d) Occupation probabilities for several lowest-in-energy states in
   the spin multiplets for $N$ and $N+1$ at $T=1.8$ K;
  (e)-(h) Transition energies $\varepsilon_{N+1}^k-\varepsilon_N^{k^\prime}$  relevant for the SET processes at the Coulomb resonance (i.e., $\varepsilon_{N+1}^0=\varepsilon_N^0$ is restored for each $B$ by tuning $V_\text{g}$) for $k,k^\prime\leqslant4$. Different colors of lines  are used to distinguish groups of transitions with respect to possible combinations of indices $k$ and $k^\prime$ (see the discussion regarding Figs.~\ref{Fig_3}(c) and~\ref{Fig_4});
  (i)-(l) energies $\varepsilon_n^k$ for $n=N,N+1$ at the Coulomb resonance -- observe that the curves for $k=0$ overlap;
  (m)-(p) Dependence of the current on the number of spin-multiplet states $r$ included from each charge state.
    The left/right most column represents the case of absent/significant transverse magnetic anisotropy.
    Importantly, each column shows a detailed analysis of selected conductance curves from Fig.~\ref{Fig_2}(c).
    We note that transition-energy lines in (e)-(h) can be easily identified with the use of \Fig{Fig_4}(a).
    It can be seen that increasing $E/D$ results in shifting the minima of the transition-energy curves in (e)-(h) towards higher values of the field.
    Such a behavior, in turn,  affects the occupation probabilities (a)-(d), so that the probability of finding an SMM either in the ground ($k=0$) or first excited ($k=1$) state for both charge states $N$ and $N+1$ remain equal for a larger magnetic-field range (compare the outermost columns). Recall that the position of the Coulomb peak is fixed mostly by $D$, see Fig.~\ref{Fig_8}.
  }
  \label{Fig_13}
\end{figure}

\begin{figure}[p]
\vspace*{75pt}
   \includegraphics[width=0.85\textwidth]{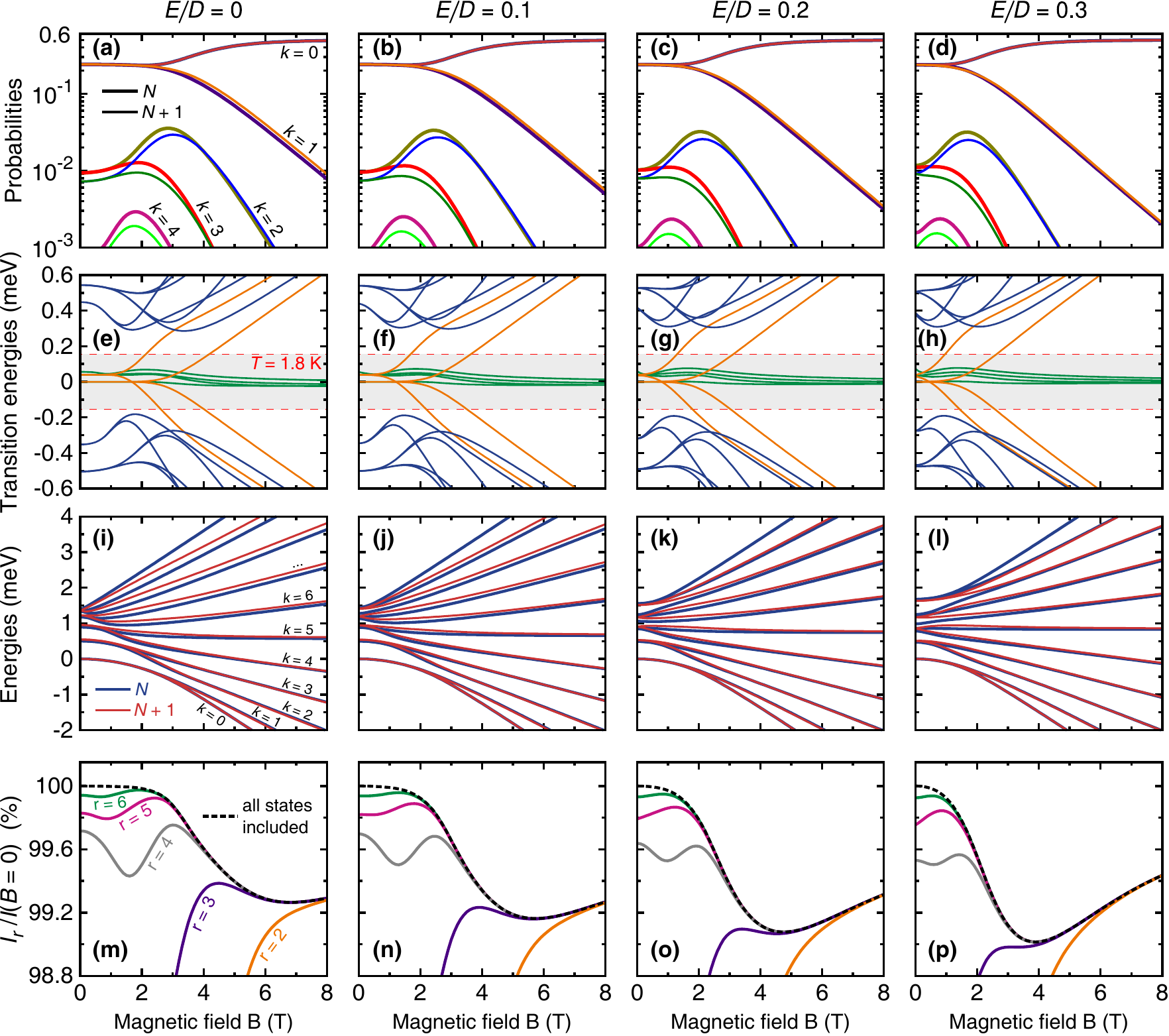}
    \caption{
    (color online)
     Signatures of the transverse magnetic anisotropy in electronic transport (magnetic field along the \emph{medium} axis, $\theta=90^\circ$ and $\phi=90^\circ$):
    Generally, this figure is analogous to \Fig{Fig_13} except that now the external magnetic field is rotated to align with the molecule's intermediate ($y$) axis.
    To begin with, we note that the results shown in the leftmost column (i.e. for $E/D=0$)  are identical to those in the leftmost column of \Fig{Fig_13}, which is the manifestation of the molecule's rotational symmetry about the easy ($z$) axis  in the absence of transverse component of magnetic anisotropy.
    Unlike for the case of $\phi=0^\circ$, the consequence of the increase of $E/D$ is the displacement of the transition-energy curves minima (e)-(h) towards smaller values of the field. Interestingly enough, in the situation under discussion one thus observes a more abrupt decrease of the current [see dashed lines in (m)-(p)] for larger $E/D$ occurring at smaller values of $B$.
  }
  \label{Fig_14}
\end{figure}

\begin{figure}[p]
\vspace*{75pt}
   \includegraphics[width=0.85\textwidth]{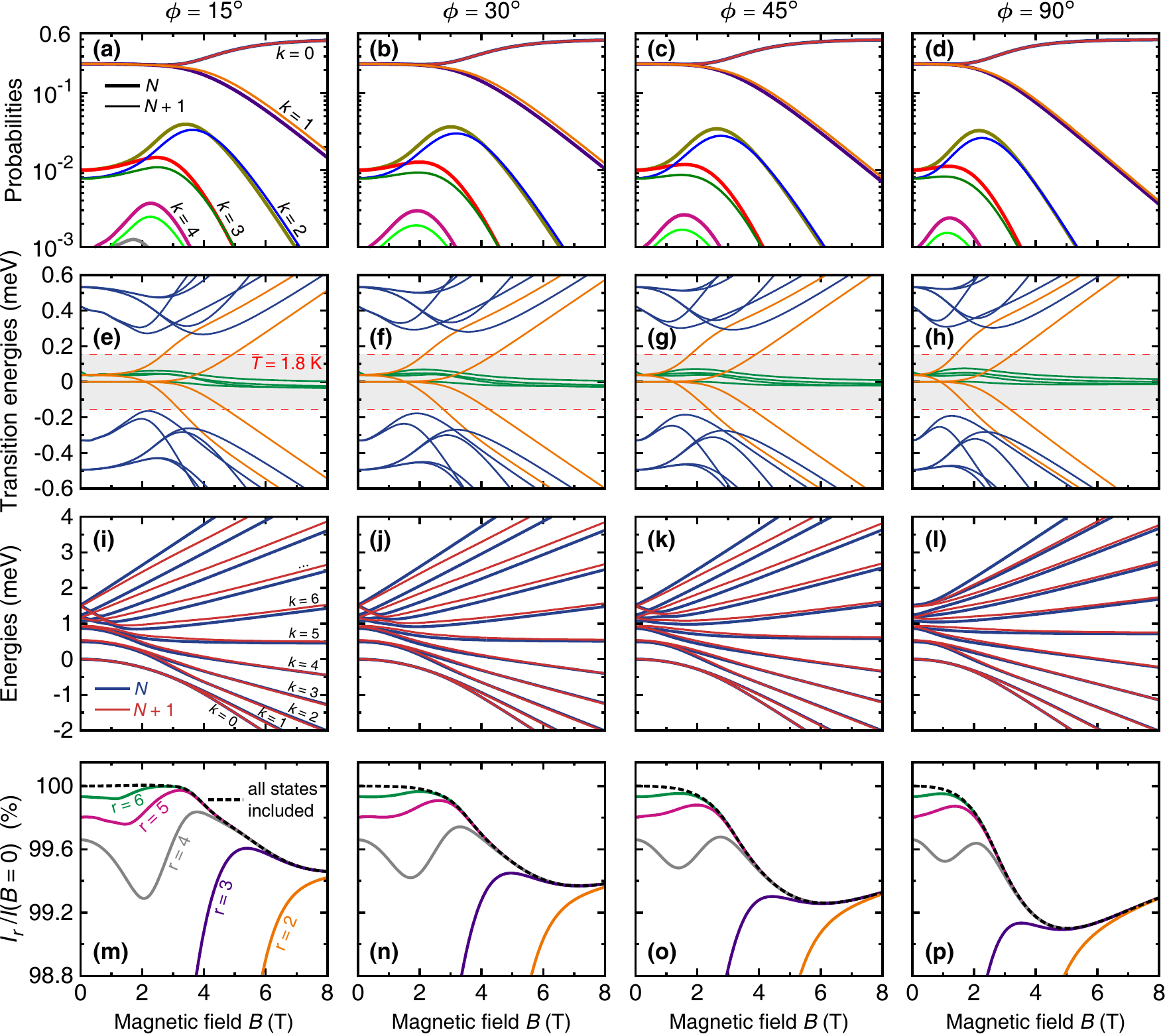}
    \caption{
    (color online)
    Dependence of  transport signatures  of the transverse anisotropy on the orientation of magnetic field in the \emph{hard} plane ($\theta=90^\circ$) for $E/D=0.17$:
     Analogous to Figs.~3(b)-(e) of the main article with each column corresponding now to a different value of $\phi$. Note that the case of $\phi=0^\circ$ is presented in Figs.~3(b)-(e) of the main article.
     Furthermore, here each column shows a detailed analysis of selected conductance curves from Fig.~2(d) of the main article.
  }
  \label{Fig_15}
\end{figure}

\begin{figure}[p]
\vspace*{50pt}
   \includegraphics[width=0.85\textwidth]{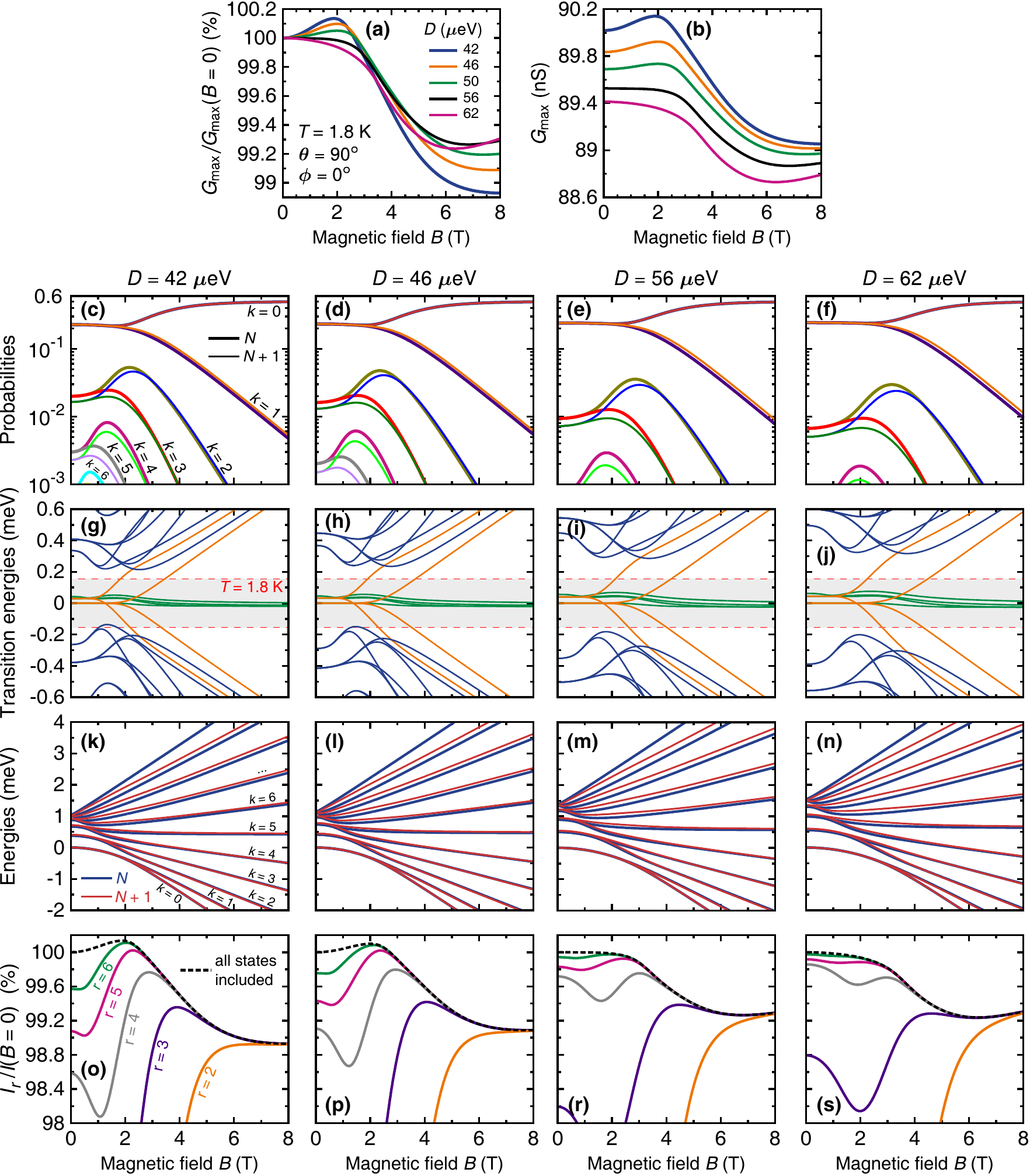}
    \caption{
    (color online)
    Evolution of the Coulomb peak amplitude in the absence of transverse magnetic anisotropy ($E=0$):
    This figure serves to illustrate the fact that even if the transverse magnetic anisotropy is absent, by making the uniaxial magnetic anisotropy parameter $D$ smaller (keeping a fixed temperature)  one can eventually also produce a maximum as for $E\neq0$.
    However, this maximum occurs at a completely different (smaller) value of magnetic field. Moreover, the shape of $G_\text{max}(B)$ remains invariant under rotation of the field in the hard plane, this is when the angle $\phi$ is varied. None of these are the case in the experiment under discussion.
    (a)-(b) Dependence of $G_\text{max}(B)$ on the value of the uniaxial magnetic anisotropy parameter $D\equiv D_N$ (and $D_{N+1}=1.2D$) for an external magnetic field applied
    along the molecule's hard axis ($\theta=90^\circ$ and $\phi=0^\circ$).
    A detailed analysis of selected curves from (a)-(b) is carried out in (c)-(s), with each column corresponding to the indicated value of $D$.
  }
  \label{Fig_16}
\end{figure}

\begin{enumerate}[(i)]
\item
Upon increase of $E$ the minima of the transition-energy curves are shifted to higher field values
 and the value achieved at the minimum is lowered, cf. Fig.~\ref{Fig_3}(c)  with Fig.~\ref{Fig_13}(e)-(h).
 For a fixed temperature, this leads to a more pronounced maximum conductance attained at a higher field value.
\item
Generally, the transition energies in Fig.~\ref{Fig_3}(c) show sharp features (i.e., oscillations below $B=2$ T) due to Berry-phase interference on which several techniques for extracting $E$ rely --  by analyzing the field dependence of the tunnel splitting between two selected states.~\cite{Gatteschi_book,Gatteschi_Angew.Chem.Int.Ed.42/2003,Wernsdorfer_Science284/1999,Burzuri_Phys.Rev.Lett.111/2013} However, the detection of such behavior in the conductance  requires very specific low temperature conditions. This is in contrast to the present experimental conditions where these Berry-phase features are averaged out when taking into account multiple accessible states.
This leaves only the large scale, collective variations of the transition energy spectrum caused by $E$ which as we have shown suffice for estimation of $E$.
In Fig.~\ref{Fig_3}(d) we illustrate the importance of taking into account many excited states for both charge states to describe current correctly.
\item
Finally, Fig.~\ref{Fig_2}(c)  shows the relative CP amplitude for increasing $E/D$. A qualitative distinction from the $E\ll D$ limit is the appearance of an additional shoulder close to $B=6$ T.
It is tempting to see such a shoulder in the sample A curve of Fig.~\ref{Fig_2}(a), although the sample B curve exhibits features of similar size where it should theoretically be smooth.
In summary, the calculations certainly show that a sizeable $E$ term  leads to fingerprints in the linear conductance as clear as those for the $D$ term, even for relatively high temperatures.
\end{enumerate}

\onecolumngrid

\clearpage

\clearpage

\clearpage

\clearpage

\clearpage

\twocolumngrid


%

\end{document}